\documentclass[11pt]{article}

\usepackage{amsmath} 
\usepackage{amsthm} 
\usepackage{amssymb}	
\usepackage{graphicx} 
\usepackage{multicol} 
\usepackage{multirow}
\usepackage{color}
\usepackage[dvips,letterpaper,margin=1in,bottom=1in]{geometry}

\usepackage[utf8]{inputenc}
\usepackage[english]{babel}

\usepackage{diagbox}
\usepackage{mathtools}

\newtheorem{theorem}{Theorem}[section]

\newtheorem{lemma}[theorem]{Lemma}
\newtheorem{corollary}[theorem]{Corollary}
\newtheorem{proposition}[theorem]{Proposition}
\newtheorem{fact}[theorem]{Fact}

\newtheorem{definition}{Definition}[section]
\newtheorem{example}{Example}
\newtheorem{remark}{Remark}[section]

\newcommand{\braket}[2]{\left< #1 \vphantom{#2} \middle| #2 \vphantom{#1} \right>} 

\DeclarePairedDelimiter\rbra{\lparen}{\rparen}
\DeclarePairedDelimiter\sbra{\lbrack}{\rbrack}
\DeclarePairedDelimiter\cbra{\{}{\}}
\DeclarePairedDelimiter\abs{\lvert}{\rvert}
\DeclarePairedDelimiter\Abs{\lVert}{\rVert}
\DeclarePairedDelimiter\ceil{\lceil}{\rceil}

\DeclarePairedDelimiter\ket{\lvert}{\rangle}
\DeclarePairedDelimiter\bra{\langle}{\rvert}

\newcommand{\set}[2] {\left\{\, #1 \colon #2 \,\right\}}

\newcommand{\tr} {\operatorname{tr}}

\newcommand{\spanspace} {\operatorname{span}}

\newcommand{\FA} {\textup{FA}}
\newcommand{\BA} {\textup{BA}}
\newcommand{\QFA} {\textup{QFA}}
\newcommand{\QBA} {\textup{QBA}}
\newcommand{\RL} {\mathsf{RL}}
\newcommand{\CFL} {\mathsf{CFL}}

\usepackage{latexsym}
\usepackage{CJK}

\usepackage{enumerate}

\usepackage{algorithm}
\usepackage{algpseudocode}

\usepackage{stmaryrd}
\usepackage{booktabs}

\usepackage{hyperref}
\newcommand{\footremember}[2]{%
    \footnote{#2}
    \newcounter{#1}
    \setcounter{#1}{\value{footnote}}%
}

\usepackage{cite}
\usepackage{bbm}

\usepackage{tablefootnote}
\usepackage{threeparttable}
\usepackage{scrextend}

\usepackage{tikz}
\usetikzlibrary{positioning}

\usepackage{qcircuit}

\begin{document}

    \title{Quantum B\"{u}chi Automata}
        \author{
            Qisheng Wang 
            \footremember{1}{Qisheng Wang is with the Graduate School of Mathematics, Nagoya University, Nagoya, Japan (e-mail: \url{QishengWang1994@gmail.com}). Part of the work was done when the author was at the Department of Computer Science and Technology, Tsinghua University, Beijing, China.}
            \and Mingsheng Ying \footremember{7}{Mingsheng Ying is with the Centre for Quantum Software and Information, University of Technology Sydney, Australia. (e-mail: \url{Mingsheng.Ying@uts.edu.au}).}
        }
        \date{}
        \maketitle

    \begin{abstract}
        Quantum finite automata (QFAs)  have been extensively studied in the literature. In this paper, we define and systematically study quantum B\"uchi automata (QBAs) over infinite words to model the long-term behavior of  quantum systems, which extend QFAs. We introduce the classes of $\omega$-languages recognized by QBAs in 
        probable, almost sure, strict and non-strict threshold semantics.
        Several pumping lemmas and closure properties for QBAs are proved. Some decision problems for QBAs are    investigated. In particular, we show that there are surprisingly only at most four substantially different classes of $\omega$-languages recognized by QBAs (out of uncountably infinite). The relationship between classical $\omega$-languages and QBAs is clarified using our pumping lemmas. We also find an $\omega$-language recognized by QBAs under the almost sure semantics, which is not $\omega$-context-free. 
    \end{abstract}


    \newpage

    \tableofcontents
    \newpage
    \section{Introduction}
    
    As acceptors for infinite words (i.e., $\omega$-words), B\"{u}chi automata \cite{Buc66} are widely applied in model-checking, program analysis and verification, reasoning about infinite games and decision problems for various logics.
    Many variants of B\"{u}chi automata have been defined in the literature, with acceptance conditions different from the original one in \cite{Buc66}, e.g., Muller, Rabin and Street conditions (cf. \cite{Tho90}). More recently, probabilistic generalizations of B\"{u}chi automata and other $\omega$-automata have been systematically studied in \cite{BG05, BGB12}.
    
    In quantum computing, quantum automata over finite words were introduced more than 20 years ago and have been extensively studied since then; see for example \cite{KW97, AF98, MC00, BP02}. But quantum automata over infinite words have  only very briefly been considered in the literature. The B\"{u}chi, Street and Rabin acceptance conditions for quantum automata were defined in   \cite{RKMS+13, DB10}. The only result obtained in \cite{RKMS+13} is an example $\omega$-language accepted by a $2$-way quantum automaton but not by any $1$-way quantum automaton. 
    Furthermore, an $\omega$-language $\rbra*{a^m b}^\omega$ that can be efficiently recognized by a $1$-way quantum automaton under the almost sure semantics was given in \cite{GPA15}, and  Bhatia and Kumar \cite{BK19} examined  the relationships between quantum B\"{u}chi, Muller, Rabin and Streett automata over $\omega$-words and their closure properties. However, in our opinion, the definition of quantum B\"{u}chi automata given in \cite{RKMS+13, DB10} is problematic (see Section \ref{sec:cmp} for a detailed discussion). 
    
    The overall aim of this paper is to properly define the notion of quantum B\"{u}chi automata and systematically study their fundamental properties, with the expectation that the results obtained here can serve as the mathematical tools needed in the areas like model-checking quantum systems \cite{YF21}, semantics and verification of quantum programs \cite{Yin16},
    and analysis of quantum games \cite{EWL99, Mey99, GW07, Zha12}.
    
    \subsection{Our Contributions}
    
    In this paper, we propose a formal definition of quantum B\"uchi automata (QBAs for short). Specifically, we adopt the idea of measure-once quantum finite automata (MO-QFA) defined in \cite{MC00}, and extend it to accept $\omega$-words under a B\"uchi-like  acceptance condition (see Definition \ref{def:qba}). Roughly speaking, the quantum B\"uchi acceptance condition we employ is that there is an accepting state $\ket*{\psi}$ such that the states in the run of the quantum finite automaton can be close enough to $\ket*{\psi}$ infinitely often. This is an intuitive generalization of  B\"uchi acceptance condition to the quantum case. 
    
    Suppose $\Sigma$ is a finite alphabet. Let $\Sigma^\omega$ denote the set of $\omega$-words over $\Sigma$.
    Under the quantum B\"uchi acceptance condition, we define the characteristic function $$f_\mathcal{A}^{\QBA} \colon \Sigma^\omega \to \sbra*{0, 1}$$ of a QBA $\mathcal{A}$. Based on it, we define $\omega$-languages recognized by QBAs under several different semantics.  Specifically, we write $\mathcal{L}_{\rhd \lambda}^{\QBA}\rbra*{\mathcal{A}}$ for the $\omega$-language, i.e., the set of $\omega$-words $w$ such that $f_\mathcal{A}^{\QBA}\rbra*{w} \rhd \lambda$, where $\rhd$ is either $>$ or $\geq$, and $\lambda \in \sbra*{0, 1}$. It is called the strict (resp. non-strict) threshold semantics if $\rhd$ is $>$ (resp. $\geq$). 
 Especially, if $\rhd \lambda$ is $>0$, it is called the probable semantics, and if $\rhd \lambda$ is $=1$, it is called the almost sure semantics.
 
    \textbf{An Equivalent Definition}: For a better understanding of QBAs, we consider another definition of quantum B\"uchi acceptance condition (see Definition \ref{def:qba-alter}) that only requires the states in the run to hit the accepting subspace infinitely often. At the first glance, it is different from the original quantum B\"uchi acceptance condition because one  requires to hit a subspace while the other requires to hit a state. But surprisingly, we show that they are equivalent (see Lemma \ref{lemma:alter}). 
    As a result, one can write the characteristic function of QBAs as the supremum limit of that of MO-QFAs (see Lemma \ref{lemma:lim}). This can be seen as a quantum generalization of the well-known relationship between classical finite automata and classical B\"uchi automata (see Fact \ref{fact:fa-ba}).
    
    \textbf{Pumping Lemmas}: To clarify  the non-inclusion relationship between various classes of $\omega$-languages, we provide several pumping lemmas for QBAs: Theorem \ref{thm:pumping-lemma-helper} is an $\omega$-generalization of the pumping lemma for MO-QFAs over finite words given in \cite{MC00}. Theorem \ref{thm:pumping-lemma} are particularly more useful to exclude an $\omega$-language that cannot be recognized by any QBAs. 
    As their applications, we examine  the relationship between QBAs (see Theorem \ref{thm:almost-sure-notin-threshold}), compare QBAs with classical B\"uchi automata (see Theorem \ref{thm:RL-notin-QBA}), and prove several closure properties of QBAs (see Theorem \ref{thm:non-closure}).
    
    \textbf{Classification of Quantum $\omega$-Languages}: We compare the classes of $\omega$-languages recognized by QBAs under different threshold semantics. Let $\mathbb{L}_{\rhd \lambda}\rbra*{\QBA}$ stand for  the class of $\omega$-languages $\mathcal{L}_{\rhd \lambda}^{\QBA}\rbra*{\mathcal{A}}$ over all QBAs $\mathcal{A}$. By definition, there should be uncountably infinitely many classes of $\omega$-languages recognized by QBAs. But surprisingly, only four of them are substantially different (see Theorem \ref{thm:relation-threshold-semantics}): 
    \[
    \begin{matrix}
        \mathbb{L}_{>0}\rbra*{\QBA}, & \mathbb{L}_{>\lambda}\rbra*{\QBA}, & \mathbb{L}_{=1}\rbra*{\QBA}, & 
        \mathbb{L}_{\geq \lambda}\rbra*{\QBA},
    \end{matrix}
    \]
    where $\lambda \in \rbra*{0, 1}$ can be any constant, e.g., $\lambda = 1/2$. Furthermore, we show that $\mathbb{L}_{>0}\rbra*{\QBA} \subseteq \mathbb{L}_{>\lambda}\rbra*{\QBA}$ and $\mathbb{L}_{=1}\rbra*{\QBA} \subseteq \mathbb{L}_{\geq\lambda}\rbra*{\QBA}$, but $\mathbb{L}_{=1}\rbra*{\QBA} \not \subseteq \mathbb{L}_{>\lambda}\rbra*{\QBA}$ for any $\lambda \in [0, 1)$. An overview of the relationship between QBAs under different semantics is depicted in Figure \ref{fig:relation-wlanguage}. \begin{figure}[!htp]\centering
    \[
        \begin{matrix}
            & & \mathbb{L}_{\geq \lambda}\rbra*{\QBA} \\
            & & \rotatebox{270}{\hspace{-0.6em}$\mbox{$\supseteq$}$} \\
            \mathbb{L}_{> \lambda}\rbra*{\QBA} & $\mbox{$\not \supseteq$}$ & \mathbb{L}_{= 1}\rbra*{\QBA} \\
            \rotatebox{270}{\hspace{-0.6em}$\mbox{$\supseteq$}$}  \\
            \mathbb{L}_{> 0}\rbra*{\QBA} 
        \end{matrix}
    \]
    \caption{Expressiveness of QBAs under different semantics.}\label{fig:relation-wlanguage}
    \end{figure}
    
    \textbf{Relationship with Classical $\omega$-Languages}: Using our pumping lemmas, we show that the $\omega$-languages recognized by classical B\"uchi automata and QBAs are  incomparable. Specifically, we show that there is an $\omega$-regular language and an $\omega$-context-free (but not regular) language that are not in $\mathbb{L}_{>\lambda}\rbra*{\QBA}$ (see Theorem \ref{thm:RL-notin-QBA}). Conversely, we show that there is an $\omega$-language in $\mathbb{L}_{>\lambda}\rbra*{\QBA}$ that is not $\omega$-regular, and there is an $\omega$-language in $\mathbb{L}_{=1}\rbra*{\QBA}$ that is not even $\omega$-context-free (see Theorem \ref{thm:QBA-notin-RL}).
    
    \textbf{Closure Properties}: Using the basic operations of QBAs, 
    we show that $\mathbb{L}_{>0}\rbra*{\QBA}$ is closed under union, and $\mathbb{L}_{>\lambda}\rbra*{\QBA}$ for $\lambda \in \rbra*{0, 1}$ is closed under union in the limit (see Theorem \ref{thm:closure-union}). However, $\mathbb{L}_{>\lambda}\rbra*{\QBA}$ for $\lambda \in [0, 1)$ is not closed under intersection or complementation (see Theorem \ref{thm:non-closure}). 
    Our results on closure properties for QBAs are summarized in Table \ref{tab:closure}.

    \begin{table*}[!htp]
    \centering
    \caption{Closure properties for QBAs.}
    \label{tab:closure}
    \begin{tabular}{cccc}
    \toprule
    Class of Languages & Operation & Closed or Not & Theorem   \\ \midrule
    $\mathbb{L}_{>0}\rbra*{\QBA}$                & Union          & Y    & Theorem \ref{thm:closure-union} (1) \\ \addlinespace
    $\mathbb{L}_{> \lambda}\rbra*{\QBA}$, $\lambda \in \rbra{0, 1}$                & Union          & Y (in the limit)    & Theorem \ref{thm:closure-union} (2) \\ \addlinespace
    $\mathbb{L}_{> \lambda}\rbra*{\QBA}$, $\lambda \in [0, 1)$                & Intersection          & N    & Theorem \ref{thm:non-closure} (1) \\ \addlinespace
    $\mathbb{L}_{> \lambda}\rbra*{\QBA}$, $\lambda \in [0, 1)$                & Complementation          & N    & Theorem \ref{thm:non-closure} (1) \\ \addlinespace
    $\mathbb{L}_{> \lambda}\rbra*{\QBA}$, $\lambda \in (0, 1)$                & Limits          & N    & Theorem \ref{thm:non-closure} (2) \\ \bottomrule
    \end{tabular}
    \end{table*}
    
    \textbf{Decision Problems}: We prove that the emptiness problem of QBAs is decidable under all semantics considered in this paper: probable, almost sure, strict and non-strict threshold semantics (see Theorem \ref{thm:empty-strict-threshold} and Theorem \ref{thm:empty-non-strict-threshold}). We also show that the emptiness problem of the intersection of two QBAs is decidable (see Theorem \ref{thm:empty-intersect}). Our results on decision problems for QBAs are summarized in Table \ref{tab:decidable-problem}.
    
    \begin{table*}[!htp]
    \centering
    \caption{Decidable problems for QBAs.}
    \label{tab:decidable-problem}
    \begin{tabular}{cccc}
    \toprule
    Decision Problem & Constraints & Decidability & Theorem   \\ \midrule
    $\mathcal{L}_{>\lambda}^{\QBA}\rbra*{\mathcal{A}} = \emptyset$                & $\lambda \in [0, 1)$          & Decidable    & Theorem \ref{thm:empty-strict-threshold} \\ \addlinespace
    $\mathcal{L}_{\geq \lambda}^{\QBA}\rbra*{\mathcal{A}} = \emptyset$                & $\lambda \in (0, 1]$          & Decidable    & Theorem \ref{thm:empty-non-strict-threshold} \\ \addlinespace
    $\mathcal{L}_{>\lambda}^{\QBA}\rbra*{\mathcal{A}} \cap \mathcal{L}_{>\lambda}^{\QBA}\rbra*{\mathcal{B}} = \emptyset$                & $\lambda \in [0, 1)$          & Decidable    & Theorem \ref{thm:empty-intersect} \\ \bottomrule
    \end{tabular}
    \end{table*}

    \subsection{Organization of This Paper}
    
    In Section \ref{sec:preliminaries}, we  review the basic notions of words, languages, automata, and their quantum generalizations. In Section \ref{sec:def-qba}, we give a formal definition of QBAs. In Section \ref{sec:basic-qba}, we present some basic properties of QBAs. Pumping lemmas for QBAs are proved in Section \ref{sec:pumping-lemmas}. The relationship between classical $\omega$-languages and QBAs is clarified in Section \ref{sec:relationship-classical-qba}. Closure properties of QBAs are studied in Section \ref{sec:closure}. Decision problems of QBAs are solved in Section \ref{sec:decision}. Finally, a brief conclusion is drawn in Section \ref{sec:discussion}.
    
    \textit{For readability, some proofs in this paper are put into the Appendices.}
    
    \section{Preliminaries} \label{sec:preliminaries}
    
    In this section, we briefly review the basic notions needed in this paper, including  $\omega$-languages, B\"uchi automata, and quantum finite automata. 
    
    \subsection{Languages and \texorpdfstring{$\omega$}{ω}-Languages}
    
    Let $\Sigma$ be a finite alphabet. A finite word $w$ over alphabet $\Sigma$ of length $\abs*{w} = n$ can be seen as a function $w \colon \sbra*{n} \to \Sigma$, where $\sbra*{n} = \cbra*{1, 2, \dots, n}$. We write $w\sbra*{i}$ to indicate the $i$-th character of $w$ for $i \in \sbra*{n}$. Especially, we write $\epsilon$ to denote the empty string with length $0$. We write $\Sigma^n$ for the set of all  words of length $n$,  $\Sigma^*$ for the set of all finite words, and  $\Sigma^+$ for the set of all non-empty finite words. A (finite) language $L$ over alphabet $\Sigma$ is a set of finite words; that is, $L \subseteq \Sigma^*$.
    
    An infinite word (a.k.a. $\omega$-word) $w$ over alphabet $\Sigma$ can be seen as a function $w \colon \mathbb{N} \to \Sigma$. We also write $w\sbra*{i}$ to indicate the $i$-th character of $w$. We write $\Sigma^\omega$ to denote the set of all $\omega$-words. An infinite language (a.k.a. $\omega$-language) $L$ over alphabet $\Sigma$ is a set of $\omega$-words; that is, $L \subseteq \Sigma^\omega$. 
    We write $w_n$ for  the prefix finite word of $w$ of length $n$, i.e., $w_n \sbra*{i} = w \sbra*{i}$ for every $i \in \sbra*{n}$.
    For any character $\sigma \in \Sigma$, we write $\sigma^\omega$ to denote the $\omega$-word that consists of the only character $\sigma$, i.e., $\sigma^\omega \sbra*{i} = \sigma$ for every $i \in \mathbb{N}$. For any non-empty finite word $u \in \Sigma^+$, we write $u^\omega$ for the $\omega$-word that repeats $u$ infinitely, i.e., $u^\omega\sbra*{i} = u\sbra*{\rbra*{i - 1} \bmod \abs*{u} + 1}$ for every $i \in \mathbb{N}$.
    
    \subsection{Finite Automata and B\"uchi Automata}
    
    Now let us recall the notions of (classical) finite  automata and (classical) B\"uchi automata. 
    
    \begin{definition} [Finite automaton \cite{HMU06}]
        A finite automaton (FA) is a tuple $\mathcal{A} = \rbra*{Q, q_0, \Sigma, \delta, F}$, where
        \begin{enumerate}
            \item $Q$ is a finite set of states;
            \item $q_0 \in Q$ is the initial state;
            \item $\Sigma$ is a finite alphabet;
            \item $\delta \colon Q \times \Sigma \to 2^{Q}$ is the transition function;
            \item $F \subseteq Q$ is the set of accepting states. 
        \end{enumerate}
    \end{definition}
    A finite automaton $\mathcal{A}$ is called deterministic if $\abs*{\delta\rbra*{q, \sigma}} = 1$ for every state $q \in Q$ and character $\sigma \in \Sigma$; otherwise, it is called nondeterministic.
    
    For every finite word $w \in \Sigma^*$, a run on input $w$ of finite automaton $\mathcal{A}$ is a finite sequence $s \colon s_0, s_1, \dots, s_{\abs*{w}}$ such that 
    \begin{enumerate}
        \item $s_0 = q_0$;
        \item $s_i \in \delta\rbra*{s_{i-1}, w\sbra*{i}}$ for every $i \in \sbra*{\abs*{w}}$.
    \end{enumerate}
    A finite word $w$ over $\Sigma$ is accepted by finite automaton $\mathcal{A}$ if there is a run $s$ on input $w$ of  $\mathcal{A}$ such that $s_{\abs*{w}} \in F$. Let $\chi_{\mathcal{A}}^{\FA} \colon \Sigma^* \to \cbra*{0, 1}$ be the characteristic function of the language accepted by finite automaton $\mathcal{A}$, i.e., $\chi_{\mathcal{A}}^{\FA} \rbra*{w} = 1$ if $\mathcal{A}$ accepts $w$ and $0$ otherwise.
    The language recognized by  $\mathcal{A}$ is defined by $$\mathcal{L}^{\FA}\rbra*{\mathcal{A}} = \set{w \in \Sigma^*}{\chi_{\mathcal{A}}^{\FA} \rbra*{w} = 1}.$$
    We write $$\mathbb{L}\rbra*{\FA} = \set{\mathcal{L}^{\FA}\rbra*{\mathcal{A}}}{\mathcal{A}\text{ is a finite automaton}}$$ to denote the class of all languages that can be recognized by a finite automaton. It is widely known that $\mathbb{L}\rbra*{\FA} = \RL$ (cf. \cite{HMU06}), where $\RL$ is the class of regular languages. 
    
    When a finite automaton is used to recognize $\omega$-words, it is usually called a B\"uchi automaton (BA). 
    For every $\omega$-word $w \in \Sigma^\omega$, 
    a run on input $w$ of BA  $\mathcal{A} = \rbra*{Q, q_0, \Sigma, \delta, F}$ is an infinite sequence $s \colon s_0, s_1, s_2, \dots$
    such that
    \begin{enumerate}
        \item $s_0 = q_0$;
        \item $s_i \in \delta\rbra*{s_{i-1}, w\sbra*{i}}$ for every $i \geq 1$.
    \end{enumerate}
    An $\omega$-word $w$ over $\Sigma$ is accepted by BA  $\mathcal{A}$ if there is a run $s$ on input $w$ such that there is an accepting state $q_f \in F$ that appears in the run $s$ for infinitely many times. That is, there is an accepting state $q_f \in F$ and an infinite increasing sequence of non-negative integers $n_1, n_2, \dots$ such that $s_{n_i} = q_f$ for every $i \geq 1$. Let $\chi_{\mathcal{A}}^{\BA} \colon \Sigma^\omega \to \cbra*{0, 1}$ be the characteristic function of the language accepted by BA  $\mathcal{A}$, i.e., $\chi_{\mathcal{A}}^{\BA} \rbra*{w} = 1$ if $\mathcal{A}$ accepts $w$ and $0$ otherwise.
    The $\omega$-language recognized by  $\mathcal{A}$ is defined by $$\mathcal{L}^{\BA}\rbra*{\mathcal{A}} = \set{w \in \Sigma^\omega}{\chi_{\mathcal{A}}^{\BA} \rbra*{w} = 1}.$$
    We write $$\mathbb{L}\rbra*{\BA} = \set{\mathcal{L}^{\BA}\rbra*{\mathcal{A}}}{\mathcal{A}\text{ is a B\"uchi automaton}}$$ to denote the class of all $\omega$-languages that can be recognized by a BA. It is known that $\mathbb{L}\rbra*{\BA} = \omega\mbox{-}\RL$ (cf. \cite{Tho90}), where $\omega\mbox{-}\RL$ is the class of $\omega$-regular languages. 
    
    The following fact reveals  the relationship between finite automata and B\"uchi automata as acceptors of finite words and $\omega$-words, respectively. 
    
    \begin{fact} \label{fact:fa-ba}
        For any finite automaton $\mathcal{A}$ and $\omega$-word $w \in \Sigma^\omega$, we have
        \[
            \chi_{\mathcal{A}}^{\BA}\rbra*{w} = \limsup_{n \to \infty} \chi_{\mathcal{A}}^{\FA}\rbra*{w_n}.
        \]
    \end{fact}
    
    \subsection{Linear Algebra for Quantum Mechanics}
    
    
    We use the notations in \cite{NC10} for quantum computation and quantum information.
    In quantum mechanics, the state of a quantum system is a unit vector $\ket{\psi}$ in a Hilbert space $\mathcal{H}$. 
    The norm (length) of a vector $\ket{\psi}$ is defined by $\Abs*{\ket{\psi}} = \sqrt{\braket{\psi}{\psi}}$, where $\braket{\psi}{\phi}$ denotes the inner product of $\ket{\psi}$ and $\ket{\phi}$. 
    For any linear operator $A$ on Hilbert space $\mathcal{H}$, the trace of $A$ is defined by $\tr\rbra*{A} = \sum_i \bra{i} A \ket{i}$, where $\cbra*{\ket{i}}$ is any orthonormal basis of $\mathcal{H}$. 
    The orthogonal complement of a subspace $V$ of Hilbert space $\mathcal{H}$ is defined by $V^{\perp} = \set{\ket{\psi} \in \mathcal{H}}{\braket{\psi}{\phi} = 0, \forall \ket{\phi} \in V}$.
    The evolution of a quantum system is described by a unitary operator $U$ such that $U^\dag U = UU^\dag = I$, where $U^\dag$ denotes the conjugate transpose of $U$ and $I$ is the identity. To retrieve information from a quantum system, we perform quantum measurements on quantum states. A quantum measurement is described by a set of (linear) operators $\cbra*{M_m}$ with the normalization condition $\sum_m M_m^\dag M_m = I$. If quantum measurement $\cbra*{M_m}$ is performed on a quantum system in state $\ket{\psi}$, then, with probability $p_m = \Abs*{M_m \ket{\psi}}^2$, the measurement outcome is $m$ and the quantum state becomes $M_m \ket{\psi} / \sqrt{p_m}$. 
    
    Let $\mathcal{H}_A = \spanspace\cbra*{\ket{i}_A}$ and $\mathcal{H}_B = \spanspace\cbra*{\ket{j}_B}$ be two Hilbert spaces, with $\cbra*{\ket{i}_A}$ and $\cbra*{\ket{j}_B}$ being the orthonormal bases of $\mathcal{H}_A$ and $\mathcal{H}_B$, respectively. The tensor product of $\mathcal{H}_A$ and $\mathcal{H}_B$ is defined by $\mathcal{H}_A \otimes \mathcal{H}_B = \spanspace\cbra*{\ket{i}_A \otimes \ket{j}_B}$.
    The direct sum of $\mathcal{H}_A$ and $\mathcal{H}_B$ is defined by $\mathcal{H}_A \oplus \mathcal{H}_B = \set{\ket{v}_A \oplus \ket{w}_B}{\ket{v}_A \in \mathcal{H}_A, \ket{w}_B \in \mathcal{H}_B}$.
    Let $W_A$ and $W_B$ be linear operators on $\mathcal{H}_A$ and $\mathcal{H}_B$, respectively. The tensor product of $W_A$ and $W_B$ is denoted by $W_A \otimes W_B$, and $\rbra*{W_A \otimes W_B} \rbra*{ \ket{i}_A \otimes \ket{j}_B } = W_A \ket{i}_A \otimes W_B \ket{j}_B$. The direct sum of $W_A$ and $W_B$ is denoted by $W_A \oplus W_B$, and $\rbra*{W_A \oplus W_B} \rbra*{\ket{i}_A \oplus \ket{j}_B} = W_A \ket{i}_A \oplus W_B \ket{j}_B$. 
    
    \subsection{Quantum Finite Automata} \label{sec:qfa}
    
    There are several different definitions of quantum finite automata in the literature. Here, we choose the most intuitive one defined in \cite{MC00},  commonly known as the MO-QFA (Measure-Once Quantum Finite Automaton). Our idea for defining  quantum B\"uchi automata is based on MO-QFA. We will use quantum finite automata (QFA) to mean MO-QFA throughout this paper unless otherwise specified. 
    
    \begin{definition} [Quantum finite automaton \cite{MC00}]
    A quantum finite automaton (QFA) is a 5-tuple $\mathcal{A} = \rbra*{\mathcal{H}, \ket*{s_0}, \Sigma, \set{ U_\sigma } { \sigma \in \Sigma }, F}$, where
    \begin{enumerate}
        \item $\mathcal{H}$ is a finite-dimensional Hilbert space;
        \item $\ket*{s_0}$ is a pure state in $\mathcal{H}$, called the initial state;
        \item $\Sigma$ is a finite alphabet;
        \item For each $\sigma \in \Sigma$, $U_\sigma$ is a unitary operator on $\mathcal{H}$;
        \item $F$ is a subspace of $\mathcal{H}$, called the space of accepting states.
    \end{enumerate}
    \end{definition}
    
    For every finite word $w \in \Sigma^*$, the run on input $w$ of QFA $\mathcal{A}$ is a  sequence $s \colon \ket*{s_0}, \ket*{s_1}, \dots, \ket{s_{\abs*{w}}}$
    of quantum states such that $\ket*{s_i} = U_{w\sbra*{i}} \ket*{s_{i-1}}$ for every $i \in \sbra*{\abs*{w}}$.
According to the Born law in quantum mechanics, a finite word $w$ over $\Sigma$ is accepted by  $\mathcal{A}$ with probability
    $$
    f_{\mathcal{A}}^{\QFA}\rbra*{w} = \Abs{P_F \ket{s_{\abs*{w}}}}^2 = \Abs{P_F U_w \ket*{s_0}}^2,$$
    where $P_F$ is the projector on subspace $F$, and $U_w = U_{w\sbra*{\abs*{w}}} \dots U_{w\sbra*{2}} U_{w\sbra*{1}}$. Then the  language recognized by QFA $\mathcal{A}$ is defined as 
    $$
    \mathcal{L}_{\rhd \lambda}^{\QFA}\rbra*{\mathcal{A}} = \set{w \in \Sigma^*}{f_{\mathcal{A}}^{\QFA} \rbra*{w} \rhd \lambda},$$
    where $\lambda \in [0, 1]$ and $\rhd \in \cbra*{>, \geq}$. Here, recognizability is treated in threshold semantics $\rhd \lambda$. Especially, if $\rhd \lambda$ is $> 0$, it is called probable semantics, and if $\rhd \lambda$ is $= 1$, it is called almost sure semantics. 
    We write
    $$
    \mathbb{L}_{\rhd \lambda}\rbra*{\QFA} = \set{\mathcal{L}_{\rhd \lambda}^{\QFA}\rbra*{\mathcal{A}}}{\mathcal{A}\text{ is a QFA}}
    $$
    to denote the class of all languages that can be recognized by a QFA under  semantics $\rhd \lambda$.
    
    \section{Definition of Quantum B\"uchi Automata} \label{sec:def-qba}
    
    Now we start to define quantum B\"uchi automata. Recall that in a classical B\"uchi automaton, a run $s$ is B\"uchi accepted if there is an accepting state $q_f \in F$ which appears in the run for infinitely many times. In the quantum case, each state $\ket*{\psi} \in \mathcal{H}$ defines a yes/no measurement $M_\psi = \cbra*{ M^\psi_\mathrm{yes}, M^\psi_\mathrm{no} },$ where $M^\psi_\mathrm{yes} = \ket*{\psi} \bra*{\psi}$ and $M^\psi_\mathrm{no} = I - M^\psi_\mathrm{yes}.$ This measurement is used to check whether the system is in the state $\ket\psi$. 
    
    \subsection{Quantum B\"uchi Acceptance}
    
    A quantum B\"uchi automaton (QBA) is a quantum finite automaton (QFA) under B\"uchi acceptance condition. In this sense, let  
    $
    \mathcal{A} = \rbra*{ \mathcal{H}, \ket{s_0}, \Sigma, \{ U_\sigma : \sigma \in \Sigma \}, F }$ be a QFA, and $w = \sigma_1 \sigma_2 \dots \in \Sigma^\omega$ an infinite word. Then the run on input $w$ of $\mathcal{A}$ is the infinite sequence of quantum states $s = \ket{s_0}, \ket{s_1}, \ket{s_2}, \dots$ with $\ket{s_n} = U_{\sigma_n}\ket{s_{n-1}}$ for all $n \geq 1$.
    
    \begin{definition} [Quantum B\"uchi acceptance condition] \label{def:qba}
    The characteristic function of QFA $\mathcal{A}$ under B\"uchi acceptance condition is defined by
        \begin{equation}\label{eq:def-qba}
            f_{\mathcal{A}}^{\QBA} (w) = \sup_{\ket{\psi} \in F} \sup_{\{ n_i \}} \inf_{i=1}^{\infty} \abs{\braket {\psi} {s_{n_i}}}^2
        \end{equation}
        for every $\omega$-word $w \in \Sigma^\omega$, 
        where $\ket{s_n}$ is the run on input $w$ of $\mathcal{A}$, $\{n_i\}$ ranges over all infinite  sequences with $0 \leq n_1 < n_2 < \dots$, and each $n_i$ is called a checkpoint.
    \end{definition}
    
    Intuitively, $\abs*{\braket {\psi} {s_{n_i}}}^2$ in Eq. (\ref{eq:def-qba}) can be understood as the similarity degree between states $\ket*{\psi}$ and $\ket*{s_{n_i}}$. The physical interpretation of the sequence $\abs*{\braket {\psi} {s_{n_i}}}^2$ for $i \geq 1$ is given by the following experiment. Take a system and perform measurement $M_\psi$ on it after it runs $n_1$ steps, $\abs*{\braket{\psi}{s_{n_1}}}^2$ is the probability that the measurement outcome is \textquotedblleft yes\textquotedblright, then discard the system. Take a second, identically prepared system, let it run $n_2$ steps, perform measurement $M_\psi$ on it, $\abs*{\braket{\psi}{s_{n_2}}}^2$ is the probability that the outcome is \textquotedblleft yes\textquotedblright, then discard the system. We can continue this procedure for an arbitrary number of steps. In fact, this procedure was often adopted by physicists in studying recurrence behaviour of quantum systems (see \cite{SJK08} for an example of quantum random walks). 
    
    In Definition \ref{def:qba}, $f_\mathcal{A}^{\QBA}\rbra*{w} \geq p$ means that there is an accepting state $\ket*{\psi} \in F$ and infinitely many states $\ket{s_{n_i}}$ in the run on input $w$ of $\mathcal{A}$ with $\abs*{\braket {\psi} {s_{n_i}}}^2 \to p$. The following proposition formally states this interpretation.
    
    \begin{proposition} \label{prop:def-qba-by-limsup}
        Let $\mathcal{A}$ be a QBA. For every $\omega$-word $w \in \Sigma^\omega$, we have
        \[
            f_{\mathcal{A}}^{\QBA} \rbra*{w} = \sup_{\ket{\psi} \in F} \limsup_{n \to \infty} \abs*{\braket{\psi}{s_n}}^2.
        \]
    \end{proposition}
    
    The following example shows how an $\omega$-word is recognized by a QBA. 
    
    \begin{example} \label{eg1} Let $\mathcal{A} = \rbra*{\mathcal{H}, \ket*{s_0}, \Sigma, \set{U_\sigma}{\sigma \in \Sigma}, F}$ be a QBA, where $\mathcal{H} = \spanspace \cbra*{ \ket0, \ket1 }$,
    $\ket{s_0} = \ket0$,
    $\Sigma = \cbra*{ a, b }$,
    $F = \spanspace \cbra*{ \ket0 }$,
    $U_a = R_x(\sqrt 2 \pi)$ and $U_b = R_x(-\sqrt 2 \pi)$.
    Here, $R_x(\cdot)$ stands for the rotation about the $x$ axe of the Bloch sphere; that is,
    $$
        R_x\rbra*{\theta} = \begin{bmatrix}
            \cos \left( \theta / 2 \right) & - \mathrm{i} \sin \left( \theta / 2 \right) \\
            - \mathrm{i} \sin \left( \theta / 2 \right) & \cos \left( \theta / 2 \right)
        \end{bmatrix}.
    $$
    The run on input $(ab)^\omega$ of $\mathcal{A}$ is $\ket{s_{2k}} = \ket 0$, and $\ket{s_{2k+1}} = \cos \left(  {\sqrt2 \pi} / 2 \right) \ket 0 - \mathrm{i} \sin \left( {\sqrt2 \pi} / 2 \right) \ket 1$
    for all $k \in \mathbb{N}$. 
    It can be seen that $f_\mathcal{A}^{\QBA}\rbra*{\rbra*{ab}^\omega} = 1$ because
    there is a sequence $n_i = 2i$ for $i \geq 1$ such that $\abs*{\braket{0}{s_{n_i}}}^2 = 1$.
    \end{example}
    
    \subsection{Comparison with Other Definitions} \label{sec:cmp}
    
    A different definition of  B\"uchi acceptance condition for quantum finite automata was introduced in \cite{RKMS+13, DB10} (see Definitions 9 and 10 in \cite{RKMS+13}). Using the notations in this paper, it can be rephrased as the following.
    
    \begin{definition} [Quantum B\"uchi acceptance condition in  \cite{RKMS+13}] \label{def:qba-err}
    The characteristic function $f_\mathcal{A}^{\textup{RK}} \colon \Sigma^\omega \to \sbra*{0, 1}$ of QFA $\mathcal{A}$ under B\"uchi acceptance condition is defined as follows: for every $\omega$-word $w \in \Sigma^\omega$, $f_\mathcal{A}^{\textup{RK}}\rbra*{w}$ is the real number $p$ such that there is an infinite  sequence $\cbra*{n_i}$ such that $0 \leq n_1 < n_2 < \dots$ and  $\Abs*{P_F \ket*{s_{n_i}}}^2 \geq p$ for all $i$. 
    \end{definition}
    
    We point out that Definition \ref{def:qba-err} is problematic with the following two reasons. Concerning these issues, we argue that our definition for quantum B\"uchi acceptance condition (Definition \ref{def:qba}) is more reasonable. 
    
    \textbf{(i) The Validity of Characteristic Function}: Suppose $f_\mathcal{A}^{\textup{RK}}\rbra*{w} = p$ for an $\omega$-word $w$ according to Definition \ref{def:qba-err}.
    Then 
    for every $0 \leq p' < p$, the acceptance condition in Definition \ref{def:qba-err} holds also for $p'$. Consequently, Definition \ref{def:qba-err} cannot uniquely determine the  value of $f_\mathcal{A}^{\textup{RK}}\rbra*{w}$.
    Indeed, this  definition is still problematic even if we define the value of $f_\mathcal{A}^{\textup{RK}}\rbra*{w}$ to be the largest $p$ that agrees with Definition \ref{def:qba-err};
    that is, 
    \begin{equation} \label{eq:qba-rk}
    \begin{aligned}
        f_\mathcal{A}^{\textup{RK}}\rbra*{w} = \max \{ p \in \sbra*{0, 1} \colon \exists\ {\rm increasing}\  \cbra*{n_i}\ {\rm s.t.}\ \forall i \geq 1,\ 
         \Abs*{P_F \ket*{s_{n_i}}}^2 \geq p \},
    \end{aligned}
    \end{equation}
    because  $f_\mathcal{A}^{\textup{RK}}\rbra*{w}$ in Eq. (\ref{eq:qba-rk})  may  not exist as shown by the following example. 
    
    \begin{example}
        Let us consider the QBA $\mathcal{A}$ in Example \ref{eg1}. The run on input $a^\omega$ of $\mathcal{A}$ is $
            \ket{s_k} = U_a^k \ket*{0} = R_x^{k} \ket*{0}
        $
        for every $k \geq 1$. It can be shown that for every $0 \leq p < 1$, there is an infinite increasing sequence $\cbra*{n_i}$ such that $\Abs*{P_F \ket*{s_{n_i}}}^2 \geq p$ for all $i \geq 1$. However, it does not hold for $p = 1$. Therefore, $f_\mathcal{A}^{\textup{RK}}\rbra*{w}$ in Eq. (\ref{eq:qba-rk})  does not exist. 
        By contrast,  $f_\mathcal{A}^{\QBA} \rbra*{a^\omega} = 1$ is well-defined with  Definition \ref{def:qba}. 
    \end{example}
    
    \textbf{(ii) The Interpretation of B\"uchi Acceptance Condition}:
    Apart from the above issues, the quantum B\"uchi acceptance in Definition \ref{def:qba-err} does not match the  definition of classical B\"uchi automata. The B\"uchi acceptance condition requires that there must be an accepting state that appears infinitely often in a run. However,  Definition \ref{def:qba-err} only requires that the states  in a run should be infinitely often close to the accepting subspace,  rather than a specific accepting state. By contrast, our definition  of quantum B\"uchi acceptance (see Eq. (\ref{eq:def-qba})) does require that the states in a run are infinitely often close to a specific accepting state in the accepting subspace. 
    
    As we will see in Subsection \ref{alternate-def}, the above two issues with Definition \ref{def:qba-err} can actually be remedied, and a modification of Definition \ref{def:qba-err} provides us with an equivalent characterization of QBAs.  

    \begin{remark} [A brief comparison with probabilistic B\"uchi automata]
        In \cite{BGB12}, the B\"uchi acceptance of probabilistic finite automata (namely, the characteristic function of probabilistic B\"uchi automata) is defined through the probability measure over all B\"uchi accepting runs, where a run is an infinite sequence of (classical) states in the automaton. 
        In comparison, the run of a QBA defined in this paper is an infinite sequence of quantum pure states in the state Hilbert space, and the characteristic function of a QBA is defined as the supreme similarity between the states in the run and the accepting states. 
    \end{remark}
    
    \subsection{Semantics of Quantum QBAs}
    
    Similar to the case of  QFAs in Section \ref{sec:qfa}, we can define the probable, almost sure, and threshold semantics for QBAs as follows. 
    
    \begin{definition} \label{def:qba-language}
        Let $\mathcal{A}$ be a QBA, $\lambda \in [0, 1]$, and $\rhd \in \{ >, \geq \}$. 
        The language recognized by $\mathcal{A}$ under  threshold semantics $\rhd \lambda$ is defined by
        \[
        \mathcal{L}_{\rhd \lambda}^{\QBA}\rbra*{\mathcal{A}} = \set{ w \in \Sigma^\omega} { f_{\mathcal{A}}^{\QBA}(w)\rhd \lambda }.
        \]
        The class of languages recognized by QBAs under threshold semantics $\rhd \lambda$ is defined by 
        \[
        \mathbb{L}_{\rhd \lambda}\rbra*{\QBA} = \set{ \mathcal{L}_{\rhd \lambda}^{\QBA}\rbra*{\mathcal{A}} } { \mathcal{A} \text{ is a QBA} }.
        \]
       In particular, if $\rhd \lambda$ is $> 0$ or $= 1$, it is called probable semantics or almost sure semantics, respectively. 
    \end{definition}
    
    Definition \ref{def:qba-language} can be seen as a quantum analog to the probable, almost sure, and threshold semantics of probabilistic $\omega$-automata defined in \cite{BG05, BGB12}.
    
    \subsection{Illustrative Examples}
    
    Let us see an example showing how the semantics define above can be actually used to describe certain behaviours of quantum systems.

    \begin{example} \label{def:illu}
        Consider a quantum system with Hilbert space $\mathcal{H}_4 = \operatorname{span}\{ \ket 0, \ket 1, \ket 2, \ket 3 \}$ and initial state $\ket 0$. It behaves as follows: repeatedly choose one of the following two unitary operators \footnote{These unitary operators were also considered in an example of  \cite{LYY14,LY14} for  reasoning  about termination  of quantum programs.}
          \[
            W_{\pm} = \frac 1 {\sqrt3} \begin{bmatrix}
                1 & 1 & 0 & \mp 1 \\
                \pm 1 & \mp 1 & \pm 1 & 0 \\
                0 & 1 & 1 & \pm 1 \\
                1 & 0 & -1 & \pm 1 \\
            \end{bmatrix}
          \]
          and apply it. 
          Our question is whether the system's state can be arbitrarily close to $\ket*{2}$ infinitely often. In other words, whether there is an $\omega$-word $w \in \cbra*{+, -}^\omega$ such that the state can be arbitrarily close to $\ket*{2}$ as we apply $W_{\pm}$ according to $w$. Formally, given QBA
          \[
          \mathcal{A} = \rbra*{ \mathcal{H}_4, \ket 0, \cbra*{ +, - }, \cbra*{ W_+, W_- }, \spanspace \cbra*{ \ket 2 } },
          \]
          the problem is to determine whether $\mathcal{L}_{=1}^{\QBA} \rbra*{\mathcal{A}} = \emptyset$?
    \end{example}
    
    The problem in the above  example is indeed the emptiness problem of QBAs under the almost sure semantics. 
    It was shown in \cite{BJKP05} that the decidability of the emptiness problem of QFAs (for finite words) depends on the specific threshold semantics; in particular, the emptiness of QFAs under almost sure semantics --- $\mathcal{L}_{=1}^{\QFA}\rbra*{\mathcal{A}} = \emptyset$? --- is undecidable. However, we will show in Section \ref{sec:decision} that the emptiness problem of QBAs is decidable under all semantics considered in this paper.
    
    \section{Basic Properties of Quantum B\"uchi Automata} \label{sec:basic-qba}
    
    In this section, we investigate basic properties of QBAs. These properties will serve as a step stone for studying the languages recognized by QBAs.
    
    \subsection{Alternative Quantum B\"uchi Acceptance}\label{alternate-def}
    
    For a better understanding of QBAs, we first give an alternative definition of quantum B\"uchi acceptance:
    
    \begin{definition} [Alternative definition of quantum B\"uchi acceptance] \label{def:qba-alter}
    The characteristic function of QFA $\mathcal{A}$ under B\"uchi acceptance condition is defined by
        \[
        f_\mathcal{A}^{\textup{alter}}\rbra*{w} = \sup_{\cbra*{n_i}} \inf_{i=1}^\infty \Abs*{ P_F \ket*{s_{n_i}} }^2,
        \]
        where $\ket*{s_n}$ is the run on input $w$ of $\mathcal{A}$, $P_F$ is the projector onto $F$, $\cbra*{ n_i }$ ranges over all infinite sequences with $0 \leq n_1 < n_2 < \dots$, and each $n_i$ is called a checkpoint.
    \end{definition}
    
    Intuitively, Definition \ref{def:qba-alter} can be understood as follows: 
    $f_\mathcal{A}^{\textup{alter}}\rbra*{w} \geq p$ means that there are infinitely many states $\ket{s_{n_i}}$ in the run on input $w$ of $\mathcal{A}$ such that $\Abs*{P_F \ket{s_{n_i}}}^2 \to p$.
    At the first glance, this  acceptance condition looks very different from Definition \ref{def:qba}, and it is hard to be regarded as a quantum counterpart of B\"{u}chi acceptance  because it only guarantees that $F$ (as a subspace) is hit infinitely often, but B\"{u}chi acceptance  requires that some specific state in $F$ is visited infinitely often. 
    But interestingly, Definitions \ref{def:qba} and \ref{def:qba-alter} are actually  equivalent; more precisely, we have:
    \begin{lemma} [Equivalent definition of QBA] \label{lemma:alter}
       For any QFA $\mathcal{A}$ and $w \in \Sigma^\omega$, it holds that
        \[
        f_\mathcal{A}^{\QBA}(w) = f_\mathcal{A}^{\textup{alter}}(w).
        \]
    \end{lemma}
    
    \begin{proof} [Proof sketch]
        The key observation is that the set of unit vectors in a finite-dimensional Hilbert space is compact. Therefore, if a run hits the accepting subspace infinitely often, then we can find a (Cauchy) subsequence of the run with a limit being a certain accepting state. 
    \end{proof}
    
    \begin{remark}
        Definition \ref{def:qba-alter} is indeed a modification of  Definition \ref{def:qba-err}. The common point is that both definitions consider the accepting subspace $F$ and measure how the state in the run falls in this subspace. The difference is that Definition \ref{def:qba-err} does not determine a unique value of the characteristic function; by contrast, Definition \ref{def:qba-alter} is well-defined by taking the supremum limit over all states in the run.
    \end{remark}

    \subsection{Relationship between QFA and QBA}
    
    We recall from  Fact \ref{fact:fa-ba} that the characteristic function of a  B\"uchi automaton is the  supremum limit of  characteristic functions of a sequence of   finite automata.
    A similar result holds for QBAs. 
    
    \begin{lemma} [QBA as a supremum limit of QFAs] \label{lemma:lim}
        For any QFA $\mathcal{A}$ and $w \in \Sigma^\omega$, it holds that
        \[
            f_\mathcal{A}^{\QBA}(w) = \limsup\limits_{n\to\infty} f_\mathcal{A}^{\QFA}(w_n).
        \]
    \end{lemma}
    \begin{proof}
        It immediately follows from Definition \ref{def:qba-alter} and Lemma \ref{lemma:alter}. 
    \end{proof}
    
    Lemma \ref{lemma:lim} establishes a connection between QFAs and QBAs. 
    As a simple application, we have:
    
    \begin{corollary} 
    \label{coro:qfaeq->qbaeq}
        Suppose $\mathcal{A}$ and $\mathcal{B}$ are two equivalent QFAs; that is, $f_\mathcal{A}^{\QFA}\rbra*{w} = f_\mathcal{B}^{\QFA}\rbra*{w}$ holds for every finite word $w \in \Sigma^*$. Then as QBAs,   $\mathcal{A}$ and $\mathcal{B}$ are also  equivalent; 
        that is, $f_\mathcal{A}^{\QBA}\rbra*{w} = f_\mathcal{B}^{\QBA}\rbra*{w}$ holds for every $\omega$-word $w \in \Sigma^\omega$.
    \end{corollary}
    
    However, 
    the following counterexample shows that the converse of Corollary \ref{coro:qfaeq->qbaeq} is not true. 
    
    \begin{example}
        Let $\mathcal{A} = \rbra*{ \mathcal{H}, \ket*{s_0}, \Sigma, \set{U_\sigma^\mathcal{A}} {\sigma \in \Sigma}, F}$ be a QBA, where $\mathcal{H} = \spanspace \cbra*{ \ket0, \ket1 }$, $\ket{s_0} = \ket0$, $\Sigma = \{ a \}$, $F = \spanspace \cbra*{ \ket0 }$, and $U_a^\mathcal{A} = R_x\rbra*{\sqrt 2 \pi}$.
        Let $\mathcal{B}$ be the same as $\mathcal{A}$ except for $U_a^\mathcal{B} = R_x\rbra*{\sqrt 3 \pi}$. It can be shown that $f_\mathcal{A}^{\QFA}(a^n) \neq f_\mathcal{B}^{\QFA}(a^n)$ for any $n \geq 1$, but $f_\mathcal{A}^{\QBA}(a^\omega) = f_\mathcal{B}^{\QBA}(a^\omega) = 1$.
    \end{example}
    
    The characteristic function of a QFA $\mathcal{A}$ can be used to  lower bound that of $\mathcal{A}$ as a QBA: 
    \begin{lemma} \label{lemma:self-rep}
        Suppose $\mathcal{A}$ is a QFA. For any $w \in \Sigma^*$ and $v \in \Sigma^+$, we have
        $f_\mathcal{A}^{\QBA}(uv^\omega) \geq f_\mathcal{A}^{\QFA}(u)$.
    \end{lemma}
    \begin{proof} [Proof sketch]
        As already noted in \cite{MC00}, the powers of any unitary operator can be arbitrarily close to the identity operator. Then we can find a sequence of suitable checkpoints during the infinite repetitions of $v$.
    \end{proof}
    
    Lemmas \ref{lemma:lim} and \ref{lemma:self-rep} will be used to prove some closure properties (see Section \ref{sec:closure}) and  pumping lemmas for QBAs (see Theorem \ref{thm:pumping-lemma}). 

    \subsection{Basic Operations of QBAs}
    
    In this subsection, we study some basic operations of QBAs and show how their characteristic functions can be derived from those of their component QBAs. 
    
Let us first recall the definitions of direct sum, tensor product and orthogonal complement of QFAs from \cite{MC00}.
    
    \begin{definition} [Operations  of QFAs \cite{MC00}] \label{def:op-qfa}
        Suppose \begin{align*}\mathcal{A}& = \rbra{ \mathcal{H}^\mathcal{A}, \ket{s_0^\mathcal{A}}, \Sigma, \set{U_\sigma^\mathcal{A}}{\sigma \in \Sigma}, F^\mathcal{A} },\\ \mathcal{B} &= \rbra{ \mathcal{H}^\mathcal{B}, \ket{s_0^\mathcal{B}}, \Sigma, \set{ U_\sigma^\mathcal{B} }{ \sigma \in \Sigma }, F^\mathcal{B} }\end{align*} are two QFAs, and $a$ and $b$ are two complex numbers with $\abs*{a}^2+\abs*{b}^2 = 1$. \begin{enumerate}\item The weighted direct sum of $\mathcal{A}$ and $\mathcal{B}$ is defined by
        \begin{align*}
        a\mathcal{A} \oplus b\mathcal{B} = ( \mathcal{H}^\mathcal{A} \oplus \mathcal{H}^\mathcal{B}, a\ket{s_0^\mathcal{A}} \oplus b\ket{s_0^\mathcal{B}}, \Sigma,
        \set{ U_\sigma^\mathcal{A} \oplus U_\sigma^\mathcal{B} } { \sigma \in \Sigma }, F^\mathcal{A} \oplus F^\mathcal{B} ).
        \end{align*}
  \item The tensor product of $\mathcal{A}$ and $\mathcal{B}$ is defined by
        \begin{align*}
        \mathcal{A} \otimes \mathcal{B} = ( \mathcal{H}^\mathcal{A} \otimes \mathcal{H}^\mathcal{B}, \ket{s_0^\mathcal{A}} \otimes \ket{s_0^\mathcal{B}}, \Sigma, 
        \set{ U_\sigma^\mathcal{A} \otimes U_\sigma^\mathcal{B} } { \sigma \in \Sigma }, F^\mathcal{A} \otimes F^\mathcal{B} ).
        \end{align*}
  \item  The orthogonal complement of $\mathcal{A}$ is defined by
        \[
        \mathcal{A}^\perp = \rbra{ \mathcal{H}, \ket{s_0}, \Sigma, \set{ U_\sigma } { \sigma \in \Sigma }, F^\perp }.
        \]
        \end{enumerate}
    \end{definition}
    
   When $\mathcal{A}$ and $\mathcal{B}$ are considered as  QBAs, we have:
    
    \begin{lemma} [Operations of QBAs] \label{lemma:op}
        Let $\mathcal{A},  \mathcal{B}$ be QBAs, and $a, b$ complex numbers with $\abs*{a}^2+\abs*{b}^2 = 1$. Then, \begin{enumerate}
   \item For weighted direct sums of QBAs, we have
    \begin{align*}
         \abs{a}^2  f_\mathcal{A}^{\QBA}(w) + \abs{b}^2 f_\mathcal{B}^{\QBA}(w)  \geq
        f_{a\mathcal{A} \oplus b\mathcal{B}}^{\QBA}(w) \geq
         \max \cbra*{ \abs{a}^2 f_\mathcal{A}^{\QBA}(w), \abs{b}^2 f_\mathcal{B}^{\QBA}(w) }   \geq \frac 1 2
        f_{a\mathcal{A} \oplus b\mathcal{B}}^{\QBA}(w).
        \end{align*}
        In particular, $
        f_{a\mathcal{A} \oplus b\mathcal{A}}^{\QBA}(w) = f_\mathcal{A}^{\QBA}(w)$.
    \item For tensor products of QBAs, we have $$
        f_{\mathcal{A} \otimes \mathcal{B}}^{\QBA}(w) \leq f_\mathcal{A}^{\QBA}(w) f_\mathcal{B}^{\QBA}(w).$$
        Especially, $f_{\mathcal{A}^{\otimes k}}^{\QBA}(w) = \rbra*{ f_\mathcal{A}^{\QBA}(w) }^k$.
        Here, $\mathcal{A}^{\otimes k}$ stands for the tensor product of $k$ copies of $\mathcal{A}$. That is, $\mathcal{A}^{\otimes k} = \mathcal{A}^{\otimes \rbra*{k-1}} \otimes \mathcal{A}$ and $\mathcal{A}^{\otimes 1} = \mathcal{A}$.
  \item For orthogonal complements of QBAs, we have $$f_\mathcal{A}^{\QBA}(w) + f_{\mathcal{A}^\perp}^{\QBA}(w) \geq 1,$$
        and the equality holds if and only if $\lim\limits_{n \to \infty} f_\mathcal{A}^{\QFA}(w_n)$ exists, where $w_n$ is the prefix of $w$ of length $n$.
      
        \end{enumerate}
    \end{lemma}

       \begin{remark}For comparison, some stronger results for QFAs were shown in \cite{MC00}: for every finite word $w \in \Sigma^*$, we have 
    \begin{equation} \label{eq:qfa-direct-sum}
    f_{a\mathcal{A} \oplus b\mathcal{B}}^{\QFA}\rbra*{w} = \abs*{a}^2 f_{\mathcal{A}}^{\QFA}\rbra*{w} + \abs*{b}^2 f_{\mathcal{B}}^{\QFA}\rbra*{w},
    \end{equation}
    \begin{equation} \label{eq:qfa-tensor-product}
    f_{\mathcal{A} \otimes \mathcal{B}}^{\QFA}\rbra*{w} = f_{\mathcal{A}}^{\QFA}\rbra*{w} f_{\mathcal{B}}^{\QFA}\rbra*{w},
    \end{equation}
        \begin{equation} \label{eq:qfa-ortho-complement}
    f_\mathcal{A}^{\QFA}(w) + f_{\mathcal{A}^\perp}^{\QFA}(w) = 1.
    \end{equation}\end{remark}

    The following lemma for scalar products of QBAs will be needed for clarifying the  relationship between different semantics of QBAs (see Theorem \ref{thm:relation-threshold-semantics}).
    
    \begin{lemma} \label{lemma:scaling-weighting}
        Let  $\mathcal{A}$ be a QBA and $\lambda \in [0, 1]$. Then for each of the following three functions $f_i\ (i=1,2,3)$:
        $f_1(w) = \lambda, f_2(w) =\lambda f_\mathcal{A}^{\QBA}(w), f_3(w) =\lambda f_\mathcal{A}^{\QBA}(w) + (1-\lambda)$,
        there is a QBA $\mathcal{B}$ such that  $f_\mathcal{B}^{\QBA}(w)=f_i(w)$ for every $\omega$-word $w \in \Sigma^\omega$.
    \end{lemma}

    \section{Pumping Lemmas} \label{sec:pumping-lemmas}
    
    In this section, we establish several pumping lemmas for QBAs. They will be used to show that certain language cannot be recognized by QBAs, and thus some non-inclusion and non-closure results are derived (see Theorem \ref{thm:almost-sure-notin-threshold}, Theorem \ref{thm:RL-notin-QBA}, and Theorem \ref{thm:non-closure}).
    
    Let us first present a pumping lemma for QBAs in terms of their characteristic functions.
    
    \begin{theorem}\label{thm:pumping-lemma-helper}
       Let $\mathcal{A}$ be a QBA. For any $w \in \Sigma^+$ and any $\varepsilon > 0$, there is a positive integer $k$ such that
       \[
            \abs*{f_\mathcal{A}^{\QBA}(uv) - f_\mathcal{A}^{\QBA}(uw^kv)} \leq \varepsilon
        \]
        for any $u \in \Sigma^*$ and $v \in \Sigma^\omega$.
        Moreover, if $\mathcal{A}$ is $n$-dimensional, there is a constant $c$ such that $k \leq (c\varepsilon)^{-n}$.
    \end{theorem}
    
    \begin{proof} [Proof sketch]
        The basic idea is similar to Lemma \ref{lemma:self-rep}. But in this case, checkpoints for $uw^kv$ are induced by the checkpoints for $uv$ after $k$ is chosen such that $U_w^k$ is close to the identity operator. 
    \end{proof}

    Theorem \ref{thm:pumping-lemma-helper} is an $\omega$-generalization of (and also inspired by) the pumping lemma for QFAs over finite words given in \cite{MC00}. 
    Nevertheless, the details of the proof are different.

    The following theorem is essentially a corollary of Theorem \ref{thm:pumping-lemma-helper}. But it is more convenient for excluding languages that cannot be recognized by QBAs (see the proofs of Theorem \ref{thm:almost-sure-notin-threshold}, Theorem \ref{thm:RL-notin-QBA}, and Theorem \ref{thm:non-closure}).
    
    \begin{theorem}[Pumping lemmas for QBAs] \label{thm:pumping-lemma}
    
    Let $L \in \mathbb{L}_{>\lambda}({\QBA})$ for some $\lambda \in [0, 1)$.
        \begin{enumerate}
          \item For any $w \in \Sigma^+$, $u \in \Sigma^*$ and $v \in \Sigma^\omega$,
    $uv \in L$ implies that there are infinitely many positive integers $k$ such that $uw^kv \in L$.
          \item For any $v \in L$, there are infinitely many prefixes $v_n$ of $v$ such that $v_nw^\omega \in L$ for all $w \in \Sigma^+$.
        \end{enumerate}
    \end{theorem}
    
    \begin{proof} [Proof sketch]
        The first item follows from Theorem \ref{thm:pumping-lemma-helper}, and the second item follows from Lemma \ref{lemma:self-rep}.
    \end{proof}

    \section{Relationship between Different Semantics}
    
    In this section, we examine  the relationship between the probable, almost sure and (non-)strict threshold semantics of QBAs.
    
    \subsection{Inclusion Relations}

    The following theorem shows that the specific threshold value is not sensitive for the classes of $\omega$-languages recognized by QBAs under threshold semantics.

    \begin{theorem} [Inclusion] \label{thm:relation-threshold-semantics}
        For every $\mu, \lambda \in \rbra*{0, 1}$, we have
    \begin{enumerate}
    \item $\mathbb{L}_{>0}({\QBA}) \subseteq \mathbb{L}_{>\mu}({\QBA}) = \mathbb{L}_{>\lambda}({\QBA})$.
    \item $\mathbb{L}_{=1}({\QBA}) \subseteq \mathbb{L}_{\geq \mu}({\QBA}) = \mathbb{L}_{\geq \lambda}({\QBA})$.
    \end{enumerate}
    \end{theorem}
    
    \begin{proof}
    We only prove part 1), and part 2) can be proved in a similar way. We proceed in two steps: 
    \begin{enumerate}
        \item Show that $\mathbb{L}_{>\mu} ({\QBA}) \subseteq \mathbb{L}_{>\lambda} ({\QBA})$ for every $0 < \lambda < \mu \leq 1$.
        Note that $0 < \frac \lambda \mu < 1$. For any QBA $\mathcal{A}$, by Lemma \ref{lemma:scaling-weighting}, there is a QBA $\mathcal{B}$ such that $$f_\mathcal{B}^{\QBA}(w) = \frac \lambda \mu f_\mathcal{A}^{\QBA}(w)$$ for every $w \in \Sigma^\omega$.
        Then, $f_\mathcal{A}^{\QBA}(w) > \mu$ if and only if $f_\mathcal{B}^{\QBA}(w) > \lambda$, which means
        $
            \mathcal{L}_{>\mu}^{\QBA}(\mathcal{A}) = \mathcal{L}_{>\lambda}^{\QBA}(\mathcal{B})
        $.
        As a result, $\mathbb{L}_{>\mu} ({\QBA}) \subseteq \mathbb{L}_{>\lambda} ({\QBA})$.
        
        \item Show that $\mathbb{L}_{>\lambda}({\QBA}) \subseteq \mathbb{L}_{>\mu}({\QBA})$ for every $0 \leq \lambda < \mu < 1$.
        Note that $0 < \frac {1-\mu} {1-\lambda} < 1$.
        For any QBA $\mathcal{A}$, by Lemma \ref{lemma:scaling-weighting}, there is a QBA $\mathcal{B}$ such that for any $w \in \Sigma^\omega$,
    \[
        f_\mathcal{B}^{\QBA}(w) = \frac {1-\mu} {1-\lambda} f_\mathcal{A}^{\QBA}(w) + \frac {\mu-\lambda} {1-\lambda}.
    \]
    Then $f_\mathcal{A}^{\QBA}(w) > \lambda$ if and only if $f_\mathcal{B}^{\QBA}(w) > \mu$, which means $\mathcal{L}_{>\lambda}^{\QBA}(\mathcal{A}) = \mathcal{L}_{>\mu}^{\QBA}(\mathcal{B})$.
    As a result, $\mathbb{L}_{>\lambda}({\QBA}) \subseteq \mathbb{L}_{>\mu}({\QBA})$.
    \end{enumerate}
    
    The above two steps together yield that for any $\mu, \lambda \in (0, 1)$, 
    it holds that $\mathbb{L}_{>0}({\QBA}) \subseteq \mathbb{L}_{>\mu}({\QBA}) = \mathbb{L}_{>\lambda}({\QBA})$.\end{proof}
    
    Theorem \ref{thm:relation-threshold-semantics} can be seen as a quantum generalization of the insensitivity of threshold values for the $\omega$-languages recognized by probabilistic B\"uchi automata (PBA) proved in \cite[Lemma 4.15 and Lemma 4.16]{BGB12}.
    This theorem will be used in Subsection \ref{sec:classification-qba} as a key tool for the classification of classes of $\omega$-languages recognized by QBAs. 
    
    \subsection{Non-Inclusion Relations}
    
    We can show that the class of $\omega$-language recognized by QBAs under the almost sure semantics is not included in  that under the strict threshold (therefore, also the probable) semantics. 
    
    \begin{theorem} [Non-inclusion] \label{thm:almost-sure-notin-threshold}
        For every $\lambda \in [0, 1)$, $\mathbb{L}_{=1}({\QBA}) \not \subseteq \mathbb{L}_{>\lambda}({\QBA})$.
    \end{theorem}
    \begin{proof} [Proof sketch]
        We prove the non-inclusion  by the pumping lemma for QBAs (Theorem \ref{thm:pumping-lemma}). 
        Let $\mathcal{A} = \rbra*{ \mathcal{H}, \ket{s_0}, \Sigma, \{ U_\sigma : \sigma \in \Sigma \}, F }$ be a QBA, where $\mathcal{H} = \operatorname{span} \{ \ket0, \ket1 \}$, $\ket{s_0} = \ket0$, $\Sigma = \{ a, b \}$, $F = \operatorname{span} \{ \ket0 \}$, and $U_a = R_x\rbra*{\sqrt 2 \pi}$ and $U_b = R_x\rbra*{-\sqrt 2 \pi}$.
    Let $L = \mathcal{L}_{=1}^{\QBA}(\mathcal{A})$.
    We use Theorem \ref{thm:pumping-lemma} (2) to show that $L \notin \mathbb{L}_{>\lambda}(\QBA)$ for any $\lambda \in [0, 1)$. To this end, 
    let us choose $v = a^\omega \in L$. For any prefix $x$ of $a^\omega$, say $x = a^n$ for some $n > 0$, we choose $w = ab \in \Sigma^+$. Note that $xw^\omega = a^n(ab)^\omega \notin L$, and thus $L \notin \mathbb{L}^{>\lambda}({\QBA})$. As a result, we have: $\mathbb{L}_{=1}({\QBA}) \not \subseteq \mathbb{L}_{>\lambda}({\QBA})$ for any $\lambda \in [0, 1)$.
    \end{proof}

    \subsection{Classification of Classes of \texorpdfstring{$\omega$}{ω}-Languages} \label{sec:classification-qba}
    
    Now by Theorem \ref{thm:relation-threshold-semantics} and Theorem \ref{thm:almost-sure-notin-threshold}, we conclude that there are at most $4$ (and at least $2$) different classes of languages defined by QBAs: 
    \[
    \begin{matrix}
        \mathbb{L}_{>0}\rbra*{\QBA}, & \mathbb{L}_{>\lambda}\rbra*{\QBA}, & \mathbb{L}_{=1}\rbra*{\QBA}, & 
        \mathbb{L}_{\geq \lambda}\rbra*{\QBA},
    \end{matrix}
    \]
    where $\lambda \in \rbra*{0, 1}$ is arbitrary. 
    
    It was shown in Theorem \ref{thm:almost-sure-notin-threshold} that $\mathbb{L}_{=1}({\QBA})$ is not included in $\mathbb{L}_{>\lambda}({\QBA})$. However, it is still not clear whether the inclusion $\mathbb{L}_{>0}({\QBA}) \subseteq \mathbb{L}_{>\lambda}({\QBA})$ and $\mathbb{L}_{=1}({\QBA}) \subseteq \mathbb{L}_{\geq \lambda}({\QBA})$ for $\lambda \in \rbra*{0, 1}$ given in Theorem \ref{thm:relation-threshold-semantics} are proper.
    
    \section{Relationship with Classical \texorpdfstring{$\omega$}{ω}-Languages} \label{sec:relationship-classical-qba}
    
    The aim of this section is to clarify the relationship between classical $\omega$-languages and QBAs. 
    
    \subsection{\texorpdfstring{$\omega$}{ω}-Languages Recognized by QBAs beyond Classical}
    
    First, we are going to show that 
    the expressive power of QBAs can go beyond $\omega$-regular languages and even $\omega$-context-free languages. 
    For this purpose, we need two more pumping lemmas for classical B\"{u}chi automata. The first is a pumping lemma for $\omega$-regular languages from \cite{CF11}: 
    
    \begin{theorem}[A pumping lemma for $\omega\mbox{-}\RL$ \cite{CF11}] \label{thm:pumping-lemma-w-reg}
        Let $L \subseteq \Sigma^\omega$ be an $\omega$-regular language. There exists an integer $n_0$ such that for any word $w \in L$, and for any integer $n \geq n_0$, $w$ can be written as $w = xyz$, where $x \in \Sigma^*, y \in \Sigma^+$, and $z \in \Sigma^\omega$, such that 
        \begin{enumerate}
            \item $\abs{x} = n$,
            \item $\abs{y} \leq n_0$, and
            \item For all $k \in \mathbb{N}$, $xy^kz \in L$.
        \end{enumerate}
    \end{theorem}
    
    The second is a pumping lemma for $\omega$-context-free languages. To the best of our knowledge, it seems new, and is of independent interest: 
    
    \begin{theorem} [A pumping lemma for $\omega\mbox{-}\CFL$] \label{thm:pumping-lemma-w-cfl}
        Let $L \subseteq \Sigma^\omega$ be an $\omega$-context-free language. Then there exists an positive integer $n$ such that each $z \in L$ can be written as $z = uvwxy$, where $u, v, w, x \in \Sigma^*$ and $y \in \Sigma^\omega$, such that 
        \begin{enumerate}
            \item $\abs{vwx} \leq n$,
            \item $\abs{vx} \geq 1$, and
            \item For all $k \in \mathbb{N}$, $uv^kwx^ky \in L$.
        \end{enumerate}
    \end{theorem}
    
    \begin{proof} [Proof sketch]
        The idea is to extend the pumping lemma for $\mathsf{CFL}$ given in \cite{HMU06} to $\omega\mbox{-}\CFL$ through its representation by the $\omega$-Kleene closure (cf. \cite{CG77a,CG77b,Lin76}).
    \end{proof}
    
    Equipped with the above pumping lemmas, we are able to show that some  $\omega$-languages  recognized by QBAs are neither  $\omega$-regular nor  $\omega$-context-free. 

    \begin{theorem} \label{thm:QBA-notin-RL}
    We have
    \begin{enumerate}
        \item $\mathbb{L}_{>\lambda}\rbra*{\QBA} \not \subseteq \omega\mbox{-}\RL$ for $\lambda \in \rbra*{0, 1}$.
        \item $\mathbb{L}_{=1}\rbra*{\QBA} \not \subseteq \omega\mbox{-}\CFL$.
    \end{enumerate}
    \end{theorem}
    
    \begin{proof} [Proof sketch]
        To prove 1), 
        let $\mathcal{A} = (\mathcal{H}, \ket{s_0}, \Sigma, \{ U_\sigma : \sigma \in \Sigma \}, F )$ be a QBA, where $\mathcal{H} = \spanspace \{ \ket0, \ket1 \}$, $\ket{s_0} = \ket0$, $\Sigma = \{ a, b \}$, $F = \spanspace \{ \ket0 \}$,  $U_a = R_x\rbra*{\sqrt 2 \pi}$ and $U_b = R_x\rbra*{-\sqrt 2 \pi}$.
        We choose $\lambda = 0.9$ and let $L = \mathcal{L}_{>\lambda}^{\QBA}(\mathcal{A})$. We use Theorem \ref{thm:pumping-lemma-w-reg} to show that $L$ is not $\omega$-regular.
        For any positive integer $n_0$, we choose the infinite word $w = a^{2n_0}b^{2n_0}(ab)^\omega \in L$ and set $n = n_0$. Then for any split $w = xyz$, where $\abs{x} = n_0$ and $1 \leq \abs{y} \leq n_0$, we can find certain non-negative integer $k$, such that  $$xy^kz = a^{2n_0+(k-1)\abs{y}}b^{2n_0}(ab)^\omega \notin L.$$ By Theorem \ref{thm:pumping-lemma-w-reg}, we conclude that $L$ is not $\omega$-regular.
        
        To prove 2), 
        let $\mathcal{A} = (H, \ket{s_0}, \Sigma, \{ U_\sigma : \sigma \in \Sigma \}, F )$ be a QBA, where $\mathcal{H} = \operatorname{span} \{ \ket0, \ket1 \}$, $\ket{s_0} = \ket0$, $\Sigma = \{ a, b, c \}$, $F = \operatorname{span} \{ \ket0 \}$, $U_a = R_x\rbra{\rbra*{\sqrt 2+\sqrt 3} \pi}$, $U_b = R_x\rbra*{-\sqrt 2 \pi}$ and $U_c = R_x\rbra*{-\sqrt 3\pi}$.
        Let $L = \mathcal{L}_{=1}^{\QBA}(\mathcal{A})$.
        We use Theorem \ref{thm:pumping-lemma-w-cfl} to show that $L$ is not $\omega$-context-free.
        For any positive integer $n$, we choose $z = \rbra*{a^nb^nc^n}^\omega \in L$. Then for any split $z = uvwxy$ where $u, v, w, x \in \Sigma^*$ and $y \in \Sigma^\omega$ with $\abs{vwx} \leq n$ and $\abs{vx} \geq 1$, we can find certain non-negative integer $k$ such that $uv^kwx^ky \notin L$. Thus, we conclude that $L \notin \omega\mbox{-}\CFL$.
    \end{proof}

    \subsection{Classical \texorpdfstring{$\omega$}{ω}-Languages Not Recognized by QBAs}
    
We can also prove a converse of Theorem \ref{thm:QBA-notin-RL} in the sense that there are some $\omega$-regular and $\omega$-context-free languages that cannot be recognized by QBAs. 
    
    \begin{theorem} \label{thm:RL-notin-QBA}
        For $\lambda \in [0, 1)$, we have
        \begin{enumerate}
            \item $\omega\mbox{-}\RL \not \subseteq \mathbb{L}_{>\lambda}\rbra*{\QBA}$.
            \item $\omega\mbox{-}\CFL \setminus \omega\mbox{-}\RL \not \subseteq \mathbb{L}_{>\lambda}\rbra*{\QBA}$.
        \end{enumerate}
    \end{theorem}
    
    \begin{proof}
        We will prove the theorem using the pumping lemma for QBAs (Theorem \ref{thm:pumping-lemma}). 
    
        To prove 1), 
        consider the  $\omega$-regular language $$L = (a+b)^*a^\omega.$$
        We use Theorem \ref{thm:pumping-lemma} (2) to show that $L \notin \mathbb{L}_{>\lambda}\rbra*{\QBA}$.
        Choose $v = a^\omega \in L$. Suppose that there are infinitely many $k$'s such that $a^k w^\omega \in L$ for any $w \in \Sigma^+$. Let $w = b$. Then we obtain  $a^kb^\omega \in L$, which is a contradiction. Therefore, $L \notin \mathbb{L}_{>\lambda}\rbra*{\QBA}$, and  $\omega\mbox{-}\RL \not \subseteq \mathbb{L}_{>\lambda}\rbra*{\QBA}$. 
        
       To prove 2), 
        consider the $\omega$-context-free but not $\omega$-regular language $$L = \set{ a^nb^n(a+b)^\omega }{ n \geq 1 }.$$ We use Theorem \ref{thm:pumping-lemma} (1) to show that $L \notin \mathbb{L}_{>\lambda}\rbra*{\QBA}$. Choose $a^nb^na^\omega = uv \in L$ with $n \geq 2$ and $w = a$, where $u = a^nb$ and $v = b^{n-1}a^\omega$. Suppose there are infinitely many $k$'s such that $uw^kv \in L$. Then $a^n b a^k b^{n-1} a^\omega \in L$, which is a contradiction. Therefore, $L \notin \mathbb{L}_{>\lambda}\rbra*{\QBA}$, and  $\omega\mbox{-}\CFL \setminus \omega\mbox{-}\RL \not \subseteq \mathbb{L}_{>\lambda}\rbra*{\QBA}$.
    \end{proof}
    
    \section{Closure Properties} \label{sec:closure}
    
    The aim of this section is to investigate the closure properties of the languages recognized by QBAs under Boolean operations (union, intersection, complementation) and limits.
    
    \subsection{Closure Results}
    
    
    Let us first consider closure properties under union. 
    
    \begin{theorem} [Closure of union]
    \label{thm:closure-union}
    We have
    \begin{enumerate}\item  $\mathbb{L}_{>0}\rbra*{\QBA}$ is closed under union. 
    \item For $\lambda \in (0, 1)$, $\mathbb{L}_{>\lambda}\rbra*{\QBA}$ is closed under union in the limit: if $L_1, L_2 \in \mathbb{L}_{>\lambda}\rbra*{\QBA}$, then there is a sequence $\left\{ L^{(k)} \in \mathbb{L}_{>\lambda}\rbra*{\QBA}:\ k \in \mathbb{N} \right\}$ such that $$\lim\limits_{k\to\infty} L^{(k)} = L_1 \cup L_2.$$
    \end{enumerate}\end{theorem}
    
    \begin{proof} [Proof sketch]
        Let $\mathcal{A}$ and $\mathcal{B}$ be two QBAs.
        To see that $\mathbb{L}_{>0}\rbra*{\QBA}$ is closed under union, let $$\mathcal{M} = \frac 1 {\sqrt2} \mathcal{A} \oplus \frac 1 {\sqrt2} \mathcal{B}.$$
        Then it can be shown that $\mathcal{L}_{>0}^{\QBA}\rbra*{\mathcal{M}} = \mathcal{L}_{>0}^{\QBA}\rbra*{\mathcal{A}} \cup \mathcal{L}_{>0}^{\QBA}\rbra*{\mathcal{B}}$.
        
        To prove 2), 
        let $$\mathcal{M}_k = \frac 1 {\sqrt2} \ \mathcal{A}^{\otimes k} \oplus \frac 1 {\sqrt2} \mathcal{B}^{\otimes k} .$$ Then by Theorem \ref{thm:relation-threshold-semantics} we have $L^{(k)} = \mathcal{L}_{>\lambda^k}^{\QBA}(\mathcal{M}_k) \in \mathbb{L}_{>\lambda}\rbra*{\QBA}$. Furthermore, it can be shown that when $k \to \infty$, $L^{(k)} \to \mathcal{L}_{>\lambda}^{\QBA}(\mathcal{A}) \cup \mathcal{L}_{>\lambda}^{\QBA}(\mathcal{B})$.
    \end{proof}

    \subsection{Non-Closure Results}
    
Next, we show that  $\mathbb{L}_{>\lambda}\rbra*{\QBA}$ is not closed under intersection, complementation, or limits by providing counterexamples.
    
    \begin{theorem} [Non-closure]
    \label{thm:non-closure}
    We have
    \begin{enumerate}
    \item $\mathbb{L}_{>\lambda}\rbra*{\QBA}$ is not closed under intersection or complementation for $\lambda \in [0, 1)$. 
    \item $\mathbb{L}_{>\lambda}\rbra*{\QBA}$ is not closed under limits for $\lambda \in (0, 1)$.
\end{enumerate}    \end{theorem}
    
    \begin{proof}
        
        To see that $\mathbb{L}_{>\lambda}\rbra*{\QBA}$ is not closed under complementation, 
        let $\mathcal{A} = \rbra*{ \mathcal{H}, \ket{s_0}, \Sigma, \set{ U_\sigma }{ \sigma \in \Sigma }, F }$ be a QBA, where $\mathcal{H} = \spanspace \{ \ket0, \ket1 \}$, $\ket{s_0} = \ket0$, $\Sigma = \{ a, b \}$, $F = \spanspace \{ \ket1 \}$, $U_a = I$ and $U_b = X$.
        Then it follows from Theorem \ref{thm:relation-threshold-semantics} that $L = \mathcal{L}_{>0}^{\QBA}\rbra*{\mathcal{A}} \in \mathbb{L}_{>0}\rbra*{\QBA} \subseteq \mathbb{L}_{>\lambda}\rbra*{\QBA}$. Let $\overline{L} = \Sigma^\omega \setminus L$ be the complement of $L$. Then we can  show that $\overline{L} \notin \mathbb{L}_{>\lambda}\rbra*{\QBA}$ by Theorem \ref{thm:pumping-lemma} (2). 
        Choose $v = a^\omega \notin L$, i.e., $a^\omega \in \overline{L}$. 
        For any prefixes $v_n = a^n$ of $a^\omega$ with $n \in \mathbb{N}$, we choose $w = b$. Then it can be shown that $v_n w^\omega = a^n b^\omega \in L$, i.e., $v_n w^\omega \notin \overline{L}$. By Theorem \ref{thm:pumping-lemma} (2), we know that $\overline{L} \notin \mathbb{L}_{>\lambda}\rbra*{\QBA}$.
        
        To see that $\mathbb{L}_{>\lambda}\rbra*{\QBA}$ is not closed under intersection, let $\mathcal{A} = \rbra*{ \mathcal{H}, \ket{s_0}, \Sigma, \set{ U_\sigma }{ \sigma \in \Sigma }, F }$ be a QBA, where $\mathcal{H} = \spanspace \{ \ket 0, \ket 1 \}$, $\ket{s_0} = \ket 0$, $\Sigma = \{ a, b \}$, $F = \spanspace \{ \ket 0 \}$, $U_a = X$ and $U_b = I$.
        Let $L = \mathcal{L}_{>0}^{\QBA}\rbra*{\mathcal{A}}$ and $L^\perp = \mathcal{L}_{>0}^{\QBA}\rbra*{\mathcal{A}^{\perp}}$. 
        By Theorem \ref{thm:relation-threshold-semantics}, we have $L, L^\perp \in \mathbb{L}_{>0}\rbra*{\QBA} \subseteq \mathbb{L}_{>\lambda}\rbra*{\QBA}$.
        Now we show that $L \cap L^\perp \notin \mathbb{L}_{>\lambda}\rbra*{\QBA}$ by Theorem \ref{thm:pumping-lemma} (2).
        Choose $v = a^\omega \in L \cap L^\perp$. This can be verified by $f_\mathcal{A}^{\QBA}(a^\omega) = f_{\mathcal{A}^\perp}^{\QBA}(a^\omega) = 1$. 
        For any prefixes $v_n = a^n$ of $a^\omega$ with $n \in \mathbb{N}$, we choose $w = b$. 
        It can be shown that $v_n w^\omega = a^n b^\omega \notin L \cap L^\perp$ by noting that $f_\mathcal{A}^{\QBA}(a^nb^\omega) f_{\mathcal{A}^\perp}^{\QBA}(a^nb^\omega) = 0$.
        Then by Theorem \ref{thm:pumping-lemma} (2), we know that $L \cap L^\perp \notin \mathbb{L}_{>\lambda}\rbra*{\QBA}$.
        
        To see that $\mathbb{L}_{>\lambda}\rbra*{\QBA}$ is not closed under limits for $\lambda \in (0, 1)$, 
        let $\mathcal{A} = \rbra*{ \mathcal{H}, \ket{s_0}, \Sigma, \set{ U_\sigma }{ \sigma \in \Sigma }, F }$ be a QBA, where $\mathcal{H} = \spanspace \{ \ket0, \ket1 \}$, $\ket{s_0} = \ket0$, $\Sigma = \{ a, b \}$, $F = \spanspace \{ \ket0 \}$, $U_a = R_x\rbra*{\frac {\sqrt 2} {100} \pi}$ and $U_b = R_x\rbra*{- \frac {\sqrt 2} {100} \pi}$.
        For every $k \geq 1$, let $L_k = \mathcal{L}_{>1-\frac 1 {k+10}}^{\QBA} (\mathcal{A})$.
        By Theorem \ref{thm:relation-threshold-semantics}, we have $L_k \in \mathbb{L}_{>1 - \frac{1}{k+10}}\rbra*{\QBA} = \mathbb{L}_{>\lambda}\rbra*{\QBA}$.
        It can be verified that $\lim\limits_{k \to \infty} L_k = \mathcal{L}_{=1}^{\QBA}(\mathcal{A})$.
        By an argument similar to the proof of Theorem \ref{thm:almost-sure-notin-threshold}, we can show that $\mathcal{L}_{=1}^{\QBA}(\mathcal{A}) \notin \mathbb{L}_{>\lambda}\rbra*{\QBA}$ for any $\lambda \in \rbra*{0, 1}$. Therefore, $\mathbb{L}_{>\lambda}\rbra*{\QBA}$ is not closed under limits for $\lambda \in (0, 1)$.
    \end{proof}

    \section{Decision Problems} \label{sec:decision}
    
    In this section, we consider several  decision problems about the emptiness of the $\omega$-languages recognized by QBAs.

    \subsection{Emptiness Problem under Strict Threshold Semantics}

    We first note that the emptiness of the $\omega$-languages recognized by QBAs under the strict threshold semantics is equivalent to the emptiness of the languages recognized by QFAs.

    \begin{lemma} \label{lemma:eq-qba-qfa-threshold}
        For any QFA $\mathcal{A}$ and threshold $\lambda \in [0, 1)$, $\mathcal{L}_{>\lambda}^{\QBA}\rbra*{\mathcal{A}} \neq \emptyset$ if and only if $\mathcal{L}_{>\lambda}^{\QFA}\rbra*{\mathcal{A}} \neq \emptyset$.
    \end{lemma}

    \begin{proof}
    We first show that $\mathcal{L}_{>\lambda}^{\QBA}\rbra*{\mathcal{A}} \neq \emptyset$ implies $\mathcal{L}_{>\lambda}^{\QFA}\rbra*{\mathcal{A}} \neq \emptyset$. If $\mathcal{L}_{>\lambda}^{\QBA}\rbra*{\mathcal{A}} \neq \emptyset$, then there is an $\omega$-word $w \in \Sigma^\omega$ such that $f_{\mathcal{A}}^{\QBA}\rbra*{w} > \lambda$. By Lemma \ref{lemma:lim}, it holds that  $$f_{\mathcal{A}}^{\QBA}\rbra*{w} = \limsup\limits_{n \to \infty} f_{\mathcal{A}}^{\QFA}\rbra*{w_n} > \lambda.$$ Then there is a checkpoint $n_i$ such that $f_\mathcal{A}^{\QFA}\rbra*{w_{n_i}} > \lambda$. Therefore, $w_{n_i} \in \mathcal{L}_{>\lambda}^{\QFA}$ and  $\mathcal{L}_{>\lambda}^{\QFA} \neq \emptyset$.
    
    Then, we show that $\mathcal{L}_{>\lambda}^{\QFA}\rbra*{\mathcal{A}} \neq \emptyset$ implies $\mathcal{L}_{>\lambda}^{\QBA}\rbra*{\mathcal{A}} \neq \emptyset$. If $\mathcal{L}_{>\lambda}^{\QFA}\rbra*{\mathcal{A}} \neq \emptyset$, then there is a finite word $w \in \Sigma^*$ such that $f_{\mathcal{A}}^{\QFA}\rbra*{w} > \lambda$. We choose any finite non-empty word $v \in \Sigma^+$. Then by Lemma  \ref{lemma:self-rep}, we obtain 
    \[
    f_{\mathcal{A}}^{\QBA}\rbra*{wv^\omega} \geq f_{\mathcal{A}}^{\QFA}\rbra*{w} > \lambda.
    \]
    Therefore, $wv^\omega \in \mathcal{L}_{>\lambda}^{\QBA}\rbra*{\mathcal{A}}$ and  $\mathcal{L}_{>\lambda}^{\QBA}\rbra*{\mathcal{A}} \neq \emptyset$.
    \end{proof}

    It was proved in \cite{BJKP05} that given a QFA $\mathcal{A}$ and a threshold $\lambda \in [0, 1)$, whether there exists a word $w \in \Sigma^*$ such that $f_\mathcal{A}^{\QFA}(w) > \lambda$ is decidable.
    This fact together with Lemma \ref{lemma:eq-qba-qfa-threshold} immediately yields the following  decidability result:

    \begin{theorem} [Decidability of the   emptiness under strict threshold semantics] \label{thm:empty-strict-threshold}
        For any QBA $\mathcal{A}$ and threshold $\lambda \in [0, 1)$, it is decidable whether $\mathcal{L}_{>\lambda}^{\QBA}\rbra*{\mathcal{A}} = \emptyset$.
    \end{theorem}

    \subsection{Emptiness Problem under Non-strict Threshold Semantics}

    Now we turn to deal with the emptiness problem under the non-strict threshold semantics. The proof techniques for this case are quite different from those  used in the last subsection for the strict threshold semantics.

    \begin{lemma} \label{lemma:reduce-non-strict}
        Let $\mathcal{A} = \rbra*{ \mathcal{H}, \ket{s_0}, \Sigma, \set{ U_\sigma }{ \sigma \in \Sigma }, F }$ be a QFA, and $\lambda \in (0, 1]$. Then the following two statements are equivalent:\begin{enumerate}\item  $\mathcal{L}_{\geq \lambda}^{\QBA}\rbra*{\mathcal{A}} \neq \emptyset;$ \item  There is an element $U \in \overline{\mathcal{U}}$ such that $f(U) \geq \lambda$, where $\overline{\mathcal{U}}$ is the closure of the semigroup $\mathcal{U} = \set { U_w }{ w \in \Sigma^* }$, and $f(U) = \Abs*{ P_F U \ket{s_0} }^2$.\end{enumerate}
    \end{lemma}
    
    \begin{proof}
        We first show that (1) implies (2).
        If $\mathcal{L}_{\geq\lambda}^{\QBA}(\mathcal{A}) \neq \emptyset$, then there is an $\omega$-word $w \in \Sigma^\omega$ such that $f_\mathcal{A}^{\QBA}(w) \geq \lambda$. By Lemma \ref{lemma:lim}, we have  $\limsup\limits_{n\to\infty} f_\mathcal{A}^{\QFA}(w_n) \geq \lambda$,
        which implies that there is a sequence $\{ n_k \}$ of checkpoints such that $\lim\limits_{k\to\infty} f_\mathcal{A}^{\QFA}(w_{n_k}) \geq \lambda$; that is $\lim\limits_{k\to\infty} f\rbra*{U_{n_k}} \geq \lambda$,
        where we use $U_{n_k}$ to denote $U_{w_{n_k}}$ for short.
        Since $\overline{\mathcal{U}} \subseteq \mathbb{C}^{n \times n}$ is closed, there is a subsequence $\{ n_{k_l} \}$, and an element $U \in \overline{\mathcal{U}}$ such that $\lim\limits_{l\to\infty} U_{n_{k_l}} = U$.
        Note that $f$ is continuous. Then we obtain:
        \[
            f(U) = \lim_{l\to\infty} f\rbra*{U_{n_{k_l}}} = \lim_{k\to\infty} f\rbra*{U_{n_k}} \geq \lambda.
        \]

        Then, we show that (2) implies (1).
        Suppose that there is an element $U \in \overline{\mathcal{U}}$ such that $f(U) \geq \lambda$.
        Then there is a sequence $\{ U_k \}$ such that $U_k \in \mathcal{U}$ for all $k$ and $\lim\limits_{k\to\infty} U_k = U$.
        Since $f(U)$ is continuous, we have $\lim\limits_{k\to\infty} f(U_k) = f(U) \geq \lambda$,
        that is,
        $$
        \forall \varepsilon > 0, \exists k_0 \in \mathbb{N}, \forall k>k_0, \abs*{f(U_k)-f(U)} < \varepsilon.
        $$
        In particular, we choose $\varepsilon = 1/n$, then there is a $k_n$ such that $$f\rbra*{U_{k_n}} > f(U)- \frac 1 n \geq \lambda - \frac 1 n.$$
        Note that for any $k \geq 1$, there is a finite word $u_k \in \Sigma^*$ such that $U_{u_k} = U_k$. Thus, we have $$f_\mathcal{A}^{\QFA}\rbra*{u_{k_n}} = f\rbra*{U_{k_n}} > \lambda- \frac 1 n.$$

        Now we construct a sequence $\cbra*{v_n}$ of finite words as follows.

        \begin{itemize} \item Initially, set $v_1 = u_{k_1}$ with
        $f_\mathcal{A}^{\QFA}(v_1) > \lambda-1$.

        \item For every $n \geq 1$, suppose we have chosen $v_n$ such that $f_\mathcal{A}^{\QFA}(v_n) > \lambda- 1 / n$.
        Let
        \[
        \varepsilon' = f_\mathcal{A}^{\QFA}\rbra*{u_{k_{n+1}}} - \lambda + \frac 1 {n+1} > 0,
        \]
        then by
        the pumping lemma of QFAs \cite[Theorem 6]{MC00},
        there is a $k \leq (c\varepsilon')^{-n}$ for some constant $c > 0$ such that
        \[
        f_\mathcal{A}^{\QFA}\rbra*{v_n^k u_{k_{n+1}}} \geq f_\mathcal{A}^{\QFA}\rbra*{u_{k_{n+1}}}-\varepsilon' > \lambda - \frac 1 {n+1}.
        \]
        We set $v_{n+1} = v_n^k u_{k_{n+1}}$. Then it holds that $$f_\mathcal{A}^{\QFA}\rbra*{v_{n+1}} > \lambda - \frac 1 {n+1}.$$
        \end{itemize}

        Using the above construction, we obtain a sequence $\cbra*{v_n}$ of finite words, where $v_{n+1} = v_n u_n$ for some $u_n \in \Sigma^*$. By Lemma \ref{lemma:lim}, we have 
        \begin{align*}
            f_\mathcal{A}^{\QBA}(w)
            & = \limsup_{n\to\infty} f_\mathcal{A}^{\QFA}(w_n) \geq \limsup_{n\to\infty} f_\mathcal{A}^{\QFA}(v_n)
            \geq \lambda.
        \end{align*}
        Therefore, we can find an $\omega$-word $w$ such that $w \in \mathcal{L}_{\geq\lambda}^{\QBA}(\mathcal{A})$, which implies that $\mathcal{L}_{\geq\lambda}^{\QBA}(\mathcal{A}) \neq \emptyset$.
    \end{proof}
    
    In order to reduce the emptiness problem under the non-strict threshold semantics into a first-order formula by Lemma \ref{lemma:reduce-non-strict}, we need the following algebraic result about unitary groups:
    
    \begin{lemma}[Algebraicity of unitary groups \cite{BJKP05}] \label{lemma:unitary-closure}
        Let $U_\sigma$ for $\sigma \in \Sigma$ be unitary matrices of dimension $n$. Let $\overline{\mathcal{U}}$ be the closure of the semigroup $\mathcal{U} = \{ U_w : w \in \Sigma^* \}$. Then $\overline{\mathcal{U}}$ is algebraic, and if $U_\sigma$'s have computable entries, we can compute a sequence of polynomials $f_1, \dots f_k, \dots$ such that
        \begin{enumerate}
          \item If $U \in \overline{\mathcal{U}}$, then $f_k(U) = 0$ for all $k$.
          \item There exists some $k$ such that $\overline{\mathcal{U}} = \set{ U } { f_i(U)=0, i = 1, 2, \dots, k }$.
        \end{enumerate}
    \end{lemma}

    Now we are able to show the decidability of the emptiness problem for non-strict thresholds. 

    \begin{theorem} [Decidability of the emptiness under non-strict threshold semantics] \label{thm:empty-non-strict-threshold}
        For any QBA $\mathcal{A}$ and threshold $\lambda \in (0, 1]$, it is decidable whether $\mathcal{L}_{\geq \lambda}^{\QBA}\rbra*{\mathcal{A}} = \emptyset$.
    \end{theorem}
    
    \begin{proof}
        By Lemma \ref{lemma:reduce-non-strict},
        the emptiness problem is equivalent to: whether there is an element $U \in \overline{\mathcal{U}}$ such that $f(U) \geq \lambda$, which can be written as a first-order formula
        \[
            \exists U \left[ \bigwedge_{i=1}^k (f_i(U)=0)  \land (f(U) \geq \lambda) \right],
        \]
        where $f_1, \dots, f_k$ are polynomials that can be computed by Lemma \ref{lemma:unitary-closure}, and $f$ is defined by $f(U) = \Abs*{ P_F U \ket{s_0} }^2$.
        To conclude the proof, we note that the above first-order formula is decidable using the Tarski-Seidenberg elimination method \cite{Ren92}.
    \end{proof}

    \subsection{Emptiness Problem under Intersection}

    In this subsection, we consider the emptiness of the intersection of languages recognized by two QBAs under the strict threshold semantics. The following lemma gives a necessary and sufficient condition for this emptiness in terms of the languages recognized by the corresponding QFAs.

    \begin{lemma} \label{lemma:empty-intersect}
        Let $\mathcal{A}$ and $\mathcal{B}$ be two QBAs, and $\lambda \in [0, 1)$. Then the following two statements are equivalent:\begin{enumerate}\item  $\mathcal{L}_{> \lambda}^{\QBA}\rbra*{\mathcal{A}} \cap \mathcal{L}_{> \lambda}^{\QBA}\rbra*{\mathcal{B}} \neq \emptyset$; \item There are two finite words $u, v \in \Sigma^*$ such that $f_\mathcal{A}^{\QFA}\rbra*{u} > \lambda$ and $f_\mathcal{B}^{\QFA}\rbra*{uv} > \lambda$.\end{enumerate}
    \end{lemma}
    
    \begin{proof}
        We first show that (1) implies (2). If $\mathcal{L}_{> \lambda}^{\QBA}\rbra*{\mathcal{A}} \cap \mathcal{L}_{> \lambda}^{\QBA}\rbra*{\mathcal{B}} \neq \emptyset$, then there is an $\omega$-word $w \in \Sigma^\omega$ such that $w \in \mathcal{L}_{> \lambda}^{\QBA}\rbra*{\mathcal{A}}$ and $w \in \mathcal{L}_{> \lambda}^{\QBA}\rbra*{\mathcal{B}}$. By Lemma \ref{lemma:lim}, there are two sequences of checkpoints $\cbra*{ n_i^\mathcal{A} }$ and $\cbra*{ n_i^\mathcal{B} }$ such that for all $i \in \mathbb{N}$, we have $f_\mathcal{A}^{\QFA}\rbra*{w_{n_i^\mathcal{A}}} > \lambda$, and $f_\mathcal{B}^{\QFA}\rbra*{w_{n_i^\mathcal{B}}} > \lambda$.
        Without loss of generality, we may assume that $n_1^\mathcal{A} < n_1^\mathcal{B}$. Suppose $w = \sigma_1 \sigma_2 \dots \in \Sigma^\omega$. Let $u = \sigma_1 \dots \sigma_{n_1^\mathcal{A}}$ and $v = \sigma_{n_1^\mathcal{A}+1} \dots \sigma_{n_1^\mathcal{B}}.$ It can be verified that $f_\mathcal{A}^{\QFA}(u) > \lambda$ and $f_\mathcal{B}^{\QFA}(uv) > \lambda$.

        Then, we show that (2) implies (1). If there are two finite words $u$ and $v$ such that $f_\mathcal{A}^{\QFA}\rbra*{u} > \lambda$ and $f_\mathcal{B}^{\QFA}\rbra*{uv} > \lambda$. 
        By Lemma \ref{lemma:self-rep}, we obtain $f_{\mathcal{A}}^{\QBA}\rbra*{uv^\omega} \geq f_{\mathcal{A}}^{\QFA}\rbra*{u} > \lambda$, and $f_{\mathcal{B}}^{\QBA}\rbra*{uv^\omega} = f_{\mathcal{B}}^{\QBA}\rbra*{(uv)v^\omega} \geq f_{\mathcal{B}}^{\QFA}\rbra*{uv} > \lambda$.
        The above arguments show that $w = uv^\omega \in \mathcal{L}_{> \lambda}^{\QBA}\rbra*{\mathcal{A}} \cap \mathcal{L}_{> \lambda}^{\QBA}\rbra*{\mathcal{B}}$.
    \end{proof}

    Using a technique similar to that in the proof of Theorem \ref{thm:empty-non-strict-threshold}, we can reduce the intersection emptiness problem into a decidable first-order formula and thus derive the following:

    \begin{theorem} [Decidability of the emptiness problem under intersection] \label{thm:empty-intersect}
        For any two QBAs $\mathcal{A}$ and $\mathcal{B}$, and any $\lambda \in [0, 1)$, it is decidable whether $$\mathcal{L}_{> \lambda}^{\QBA}\rbra*{\mathcal{A}} \cap \mathcal{L}_{> \lambda}^{\QBA}\rbra*{\mathcal{B}} = \emptyset.$$
    \end{theorem}
    \begin{proof}
        Let $\mathcal{M}_\mathcal{A} = 1 \mathcal{A} \oplus 0 \mathcal{B}$ and $\mathcal{M}_\mathcal{B} = 0 \mathcal{A} \oplus 1 \mathcal{B}$. 
        We note that the unitary operators of $\mathcal{M}_\mathcal{A}$ are the same as those of $\mathcal{M}_\mathcal{B}$, and write $\mathcal{U}$ for the semigroup induced by the unitary operators of $\mathcal{M}_\mathcal{A}$. By Eq. (\ref{eq:qfa-direct-sum}), we have $f_{\mathcal{M}_\mathcal{A}}^{\QFA}\rbra*{w} = f_{\mathcal{A}}^{\QFA}\rbra*{w}$ and $f_{\mathcal{M}_\mathcal{B}}^{\QFA}\rbra*{w} = f_{\mathcal{B}}^{\QFA}\rbra*{w}$ for every finite word $w \in \Sigma^*$.
        By Lemma \ref{lemma:empty-intersect}, the emptiness problem of the intersection of $\mathcal{L}_{> \lambda}^{\QBA}\rbra*{\mathcal{A}}$ and $\mathcal{L}_{> \lambda}^{\QBA}\rbra*{\mathcal{B}}$ is equivalent to:
        whether there are two elements $U, V \in \overline{\mathcal{U}}$ such that $f_\mathcal{A}(U) > \lambda$ and $f_\mathcal{B}(VU) > \lambda$,
        which can be written as a first-order formula
        \begin{align*}
        \exists U \exists V \left[ \bigwedge_{i=1}^k \rbra*{ f_i(U)=0 \land f_i(V)=0 } 
        \land \rbra*{ f_\mathcal{A}(U) > \lambda \land f_\mathcal{B}(VU) > \lambda } \right],
        \end{align*}
        where $f_1, \dots, f_k$ are polynomials that can be computed by Lemma \ref{lemma:unitary-closure}, and $f_{\mathcal{A}}$ and $f_{\mathcal{B}}$ are defined by $f_{\mathcal{A}}(U) = \Abs*{ P_{F^{\mathcal{A}}} U \ket*{s_0^{\mathcal{A}}} }^2$ and $ 
            f_{\mathcal{B}}(U) = \Abs*{ P_{F^{\mathcal{B}}} U \ket*{s_0^{\mathcal{B}}} }^2$.
        This is a first-order formula and can be decided by the Tarski-Seidenberg elimination method \cite{Ren92}.
    \end{proof}
    
    \section{Conclusion and Discussions} \label{sec:discussion}
    
    In this paper, we  defined the notion of QBAs and studied the $\omega$-languages recognized by QBAs under the probable, almost sure, (non-)strict threshold semantics.
    In particular, we established several  pumping lemmas for QBAs and use them to prove closure properties of QBAs and to clarify the relationship between the  $\omega$-languages recognized by classical B\"{u}chi automata and QBAs. 
    
    However, some basic problems are still unsolved. For example, the equivalence checking problem for QFAs has been extensively studied in the literature, e.g., \cite{Kos01,LQ06,LQ08,WLY21}. Corollary \ref{coro:qfaeq->qbaeq} enables us to derive the equivalence between two QBAs from the equivalence between them as QFAs. But there is still no algorithm for checking equivalence between QBAs.
        We leave it for future research. Another interesting open problem is  minimization of QBAs, which aims to find the minimal (possibly not unique) automaton  that has the same characteristic function over $\omega$-words as that of a given QBA. The minimization problem of QFAs over finite words has been studied in \cite{MQL12,WLY21}. But it seems that the methods developed there do not immediately apply to the case of QBAs over $\omega$-words.

    The definition of the QBA in this paper is based on and comparable to the measure-once quantum finite automaton (MO-QFA) \cite{MC00}. 
    However, other definitions of QFAs such as the measure-many quantum finite automaton (MM-QFA) \cite{KW97} were also investigated in the literature (see \cite{AY21} for a detailed review).
    It would be interesting to study how to define the B\"uchi acceptance based on other QFAs, and what surprising properties can be discovered in those cases.
    
     Except solving the  problems mentioned above, the following topics are also interesting for future research. 
    First, inspired by a recent application of QFAs in checking equivalence of sequential quantum circuits \cite{WLY22}, we would like to see  applications of QBAs in the fields like  model-checking quantum systems \cite{YF21, Yin21} as well as analysis and verification of quantum     programs \cite{Yin16}.
Second, a quantum generalization of timed  automata \cite{AD94} can be introduced based on the results obtained in this paper. We believe that such a quantum model will be very useful in not only quantum computing but also other  quantum technologies,  including real-time and embedded quantum systems.

    \section*{Acknowledgements}

    Qisheng Wang would like to thank Yangjia Li and Shenggang Ying for valuable discussions. 
    
    Qisheng Wang was also supported in part by the MEXT Quantum Leap Flagship Program (MEXT Q-LEAP) grants No. JPMXS0120319794.

    \addcontentsline{toc}{section}{References}
    
    \bibliographystyle{unsrturl}
    \bibliography{main}

    \appendix
    
    \section{Proofs for Basic Properties of QBAs}
    
    \subsection{Proof of Lemma \ref{lemma:alter}} \label{sec:proof-lemma-alter}
    
    Before the proof of Lemma \ref{lemma:alter}, we first give some equivalent interpretations of $f_\mathcal{A}^{\QBA}\rbra*{w} > \lambda$ and $f_\mathcal{A}^{\textup{alter}}\rbra*{w} > \lambda$, respectively, according to their definitions for better understanding of quantum B\"uchi acceptance condition.
    
    \begin{lemma} \label{lemma:qba-alter-interpretation}
        Let $\mathcal{A} = \rbra*{\mathcal{H}, \ket{s_0}, \Sigma, \set{ U_\sigma }{ \sigma \in \Sigma }, F }$ be a QBA, and $\lambda \in [0, 1]$ be a real number. Let $w \in \Sigma^\omega$, and let $\ket{s_n}$ be the run on input $w$ of $\mathcal{A}$. Then,
        \begin{enumerate}
          \item $f_\mathcal{A}^{\QBA}(w) > \lambda$ is equivalent to either of the following condition.
            \begin{enumerate}
                \item $\exists \varepsilon > 0, \exists \ket\psi \in F, \exists \{ n_i \}, \forall i \in \mathbb{N}, \abs*{\braket{\psi}{s_{n_i}}}^2 > \lambda+\varepsilon$.
                \item $\exists \ket\psi \in F, \exists \{ n_i \}, \lim\limits_{i\to\infty} \abs{\braket{\psi}{s_{n_i}}}^2 > \lambda.$
            \end{enumerate}
          \item $f_\mathcal{A}^{\textup{alter}}(w) > \lambda$ is equivalent to either of the following condition.
            \begin{enumerate}
                \item $\exists \varepsilon > 0, \exists \{ n_i \}, \forall i \in \mathbb{N}, \Abs*{P_F\ket{s_{n_i}}}^2 > \lambda+\varepsilon$.
                \item $\exists \{ n_i \}, \lim\limits_{i\to\infty} \Abs{P_F \ket{s_{n_i}}}^2 > \lambda$.
            \end{enumerate}
        \end{enumerate}
    \end{lemma}
    \begin{proof}
        For $f_\mathcal{A}^{\QBA}(w) > \lambda$, the interpretation to condition (a) is straightforward; and the variable $\varepsilon$ in condition (a) can be reduced to a limit form as condition (b). A similar argument also applies to $f_\mathcal{A}^{\textup{alter}}(w) > \lambda$.
    \end{proof}
    
    Now we are ready to prove Lemma \ref{lemma:alter}. The key observation is that the set of unit vectors in a finite-dimensional Hilbert space is compact. Therefore, if the run infinitely hits the accepting subspace, then we can find a subsequence of the run with a limit being a certain accepting state. 
    
    \begin{proof} [Proof of Lemma \ref{lemma:alter}]
        It is sufficient to show that for every $\lambda > 0$, $f_\mathcal{A}^{\QBA}\rbra*{w} > \lambda$ if and only if $f_\mathcal{A}^{\textup{alter}}\rbra*{w} > \lambda$.
        
        ``$\Longrightarrow$''. If $f_\mathcal{A}^{\QBA}(w) > \lambda$, by Lemma \ref{lemma:qba-alter-interpretation}, we have:
        $\exists \varepsilon > 0, \exists \ket\psi \in F, \exists \{ n_i \}, \forall i \in \mathbb{N}, \abs*{\braket{\psi}{s_{n_i}}}^2 > \lambda+\varepsilon$. 
        Let $\cbra*{\ket{v_k}}$ be the orthogonal basis of $F$ and $\ket\psi = \sum_k a_k \ket{v_k}$ with $\sum_k \abs{a_k}^2 = 1$.
        Note that
        \begin{align*}
            \abs*{ \braket{\psi}{s_{n_i}} }^2
            & = \abs*{ \sum_k a_k^* \braket{v_k}{s_{n_i}} }^2 \\
            & \leq \sum_k \abs{a_k}^2 \sum_k \abs*{\braket{v_k}{s_{n_i}}}^2 \\
            & = \sum_k \abs*{\braket{v_k}{s_{n_i}}}^2 \\
            & = \Abs*{P_F \ket{s_{n_i}}}^2.
        \end{align*}
        Therefore, $\Abs*{P_F \ket{s_{n_i}}}^2 \geq \abs*{ \braket{\psi}{s_{n_i}} }^2 > \lambda+\varepsilon$.
        By Lemma \ref{lemma:qba-alter-interpretation}, it holds that $f_\mathcal{A}^{\textup{alter}}(w) > \lambda$.

        ``$\Longleftarrow$''. If $f_\mathcal{A}^{\textup{alter}}(w) > \lambda$, by Lemma \ref{lemma:qba-alter-interpretation}, we have:
        $\exists \varepsilon > 0, \exists \{ n_i \}, \forall i \in \mathbb{N}, \Abs*{P_F\ket{s_{n_i}}}^2 > \lambda+\varepsilon$.
        Since the set of states is compact, we assert that
        $\exists \ket{\hat s} \in \mathcal{H}, \exists \{ m_i \}. \lim\limits_{i\to\infty} \ket{s_{n_{m_i}}} = \ket{\hat s}.$
        Let $\ket\psi = \frac {P_F \ket{\hat s}} {\Abs*{P_F \ket{\hat s}}} \in F$.
        Then,
        \begin{align*}
        \lim_{i\to\infty} \abs{\braket{\psi}{s_{n_{m_i}}}}^2
        & = \lim_{i\to\infty} \frac {\abs{\bra{\hat s} P_F \ket{s_{n_{m_i}}}}^2} {\Abs{P_F \ket{\hat s}}^2} \\
        & = \frac {\abs{\bra{\hat s} P_F \ket{\hat s}}^2} {\Abs{P_F \ket{\hat s}}^2} \\
        & = \Abs{P_F \ket{\hat s}}^2 \\
        & = \lim_{i\to\infty} \Abs{P_F \ket{s_{n_{m_i}}}}^2 \\
        & \geq \lambda+\varepsilon > \lambda.
        \end{align*}
    By Lemma \ref{lemma:qba-alter-interpretation}, we obtain $f_\mathcal{A}^{\QBA}(w) > \lambda$.
    \end{proof}
    
    \subsection{Proof of Lemma \ref{lemma:self-rep}} \label{sec:proof-lemma-self-rep}

    In order to prove Lemma \ref{lemma:self-rep}, we need the following two lemmas: the first one is about number theory of real numbers; and the second one is about approximation of matrices.
    
    \begin{lemma} \label{lemma:real-approx}
        For any $n$ real numbers $\alpha_1, \alpha_2, \dots, \alpha_n$, and for any $0 < \varepsilon < 1$, there are infinitely many positive integers $k$ such that
        \[
        \min \cbra*{ \{ k\alpha_j \}, \{ -k\alpha_j \} }< \varepsilon
        \]
        for all $1 \leq j \leq n$, where $\{x\}$ is the non-negative fractional part of $x$. Moreover, the smallest $k$ can be bounded by $k \leq \rbra*{c\varepsilon}^{-n}$ for some constant $c > 0$.
    \end{lemma}
    \begin{proof}
        Let $M = \ceil{1/\varepsilon}$ and $a_j = j/M$ for $0 \leq j \leq M$. Then the set $\set{ [a_{j-1}, a_j) }{ 1 \leq j \leq M }$ forms a partition of $[0, 1)$. Now that there are $M$ parts in the partition, we define a function: $h(x) = j$ if and only if $\{x\} \in [a_{j-1}, a_j)$.
        We further define: 
        \begin{align*}
            g(k) & = \rbra*{ h(k\alpha_1), h(k\alpha_2), \dots, h(k\alpha_n) } \\
            & \in \{ 1, 2, \dots, M \}^n.
        \end{align*}
        By the Pigeonhole Principle, there are two integers $1 \leq p < q \leq M^n+1$ such that $g(p) = g(q)$, i.e. $h(p\alpha_j) = h(q\alpha_j)$ for all $1 \leq j \leq n$. Then $\abs*{\{p\alpha_j\} - \{q\alpha_j\}} < 1/M$.
        Let $k = q-p \leq M^n$. Note that
        \begin{align*}
          \abs*{\{p\alpha_j\} - \{q\alpha_j\}}
          & = \min \cbra*{ \{(p-q)\alpha_j\}, \{(q-p)\alpha_j\} } \\
          & = \min \cbra*{ \{k\alpha_j\}, \{-k\alpha_j\} }.
       \end{align*}
        Then we have $\min \cbra*{ \{k\alpha_j\}, \{-k\alpha_j\} } < 1/M \leq \varepsilon$.
        On the other hand, there is a constant $c > 0$ such that $\ceil{1 / \varepsilon} \leq \rbra*{c\varepsilon}^{-1}$, e.g., $c = 1/2$. Hence, $k \leq M^n \leq \rbra*{c\varepsilon}^{-n}$.
        Moreover, since $g(k)$ can take only a finite number of different values, we can find an infinite sequence $q_1, q_2, \dots$ such that $g(q_1) = g(q_i)$ for all $i \in \mathbb{N}$. We choose $k_i = g(q_{i+1})-g(q_1)$ for all $i \in \mathbb{N}$. Then we can verify that each $k_i$ satisfies the condition $\min \cbra*{ \{ k\alpha_j \}, \{ -k\alpha_j \} } < \varepsilon$.
        \end{proof}
    
    \begin{lemma} \label{lemma:Dk-approx-for-pumping}
        Let $D$ be an $n$-dimensional unitary diagonal matrix. For every $0 < \delta < 1$, there are infinitely many positive integers $k$ such that
        \[
            D^k = I + \delta J,
        \]
        where $J$ is a diagonal matrix with $\tr\rbra*{J^\dag J} < 1$. Moreover, the smallest $k$ is bounded by $k \leq \rbra*{c\delta}^{-n}$ for some constant $c > 0$.
    \end{lemma}
    \begin{proof}
        Although this lemma was also implicitly used in the proof of the pumping lemma for QFAs over finite words in \cite[Theorem 6]{MC00}, we give a rigorous proof for it for completeness. 
        
        Assume that 
        \[
        D = \operatorname{diag}\rbra*{ \exp(\mathrm{i}\theta_1), \exp(\mathrm{i}\theta_2), \dots, \exp(\mathrm{i}\theta_n) },
        \]
        where $\theta_j = 2\alpha_j\pi$ and $\alpha_j$ is a real number for all $1 \leq j \leq n$.
        By Lemma \ref{lemma:real-approx}, there are infinitely many positive integers $k$ such that
        \[
        \min\{ \{k\alpha_j\}, \{-k\alpha_j\} \} < \frac \delta {2n\pi}
        \]
        for all $1 \leq j \leq n$,
        where the smallest $k$ can be bounded by $k \leq (c\delta)^{-n}$ for some constant $c > 0$. Let $J = (I - D^k)/\delta$, then
        \[
        J = \frac 1 \delta \operatorname{diag} \rbra*{ 1-\exp(\mathrm{i}k\theta_1), \dots, 1-\exp(\mathrm{i}k\theta_n) }.
        \]
        Note that for any $1 \leq j \leq n$,
        \[
        \begin{aligned}
        \abs*{ 1 - \exp(\mathrm{i}k\theta_j) }
        & = \sqrt{2-2\cos(k\theta_j)} \\
        & = \sqrt{2-2\cos(2k\alpha_j\pi)} \\
        & = \sqrt{2-2\cos(2 \{ k\alpha_j \} \pi)} \\
        & \leq \sqrt{2 \cdot \frac 1 2 (2\pi \min \{ \{k\alpha_j\}, \{-k\alpha_j\} \})^2} \\
        & = 2 \pi \min \{ \{k\alpha_j\}, \{-k\alpha_j\} \} \\
        & < 2 \pi \cdot \frac \delta {2n\pi} = \frac \delta n.
        \end{aligned}
        \]
        Thus, $\abs{J_{ii}} < 1/n \leq 1$ for all $1 \leq i \leq n$, and
        \[
        \tr\rbra*{J^\dag J} = \sum_{i=1}^n \abs{J_{ii}}^2 \leq \sum_{i=1}^n \abs{J_{ii}} < \sum_{i=1}^n \frac 1 n = 1.
        \]
    \end{proof}
    
    Now we are ready to prove Lemma \ref{lemma:self-rep}. 
    
    \begin{proof} [Proof of Lemma \ref{lemma:self-rep}]
        By Lemma \ref{lemma:lim}, it holds that
        \begin{align*}
            f_\mathcal{A}^{\QBA}(uv^\omega)
            & = \limsup_{n\to\infty} f_\mathcal{A}^{\QFA}\left((uv^\omega)_n\right) \\
            & \geq \limsup_{n\to\infty} f_\mathcal{A}^{\QFA}(uv^n). 
        \end{align*}
        Let $U_v = V^\dag DV$, where $V$ is a unitary matrix and $D$ is a diagonal matrix.
        By Lemma \ref{lemma:Dk-approx-for-pumping}, for any $0 < \delta < 1$, there are infinitely many integers $k$ such that
        $D^k = I+\delta J$,
        where $J$ is a diagonal matrix with
        $\tr\rbra*{J^\dag J} < 1$.
        Note that
        \begin{align*}
            f_\mathcal{A}^{\QFA}(uv^k)
            & = \Abs*{P_F U_v^k \ket{s_{\abs{u}}}}^2 \\
            & = \Abs*{P_F V^\dag D^k V \ket{s_{\abs{u}}} }^2 \\
            & = \Abs*{P_F V^\dag (I+\delta J) V \ket{s_{\abs{u}}} }^2 \\
            & = \Abs*{P_F \ket{s_{\abs{u}}} + \delta P_F V^\dag J V \ket{s_{\abs{u}}} }^2 \\
            & \geq \left( \Abs*{P_F \ket{s_{\abs{u}}}} - \delta \Abs*{P_F V^\dag J V \ket{s_{\abs{u}}}} \right)^2 \\
            & \geq \Abs*{P_F \ket{s_{\abs{u}}}}^2 - 2\delta \\
            & = f_\mathcal{A}^{\QFA}(u) - 2\delta.
        \end{align*}
        Putting this result into supremum limit, we obtain:
        \[
            \limsup_{n\to\infty} f_\mathcal{A}^{\QFA}(uv^n) \geq f_\mathcal{A}^{\QFA}(u)-2\delta.
        \]
        Since $\delta$ can be arbitrarily small, let $\delta \to 0$, it holds that
        $
            \limsup\limits_{n\to\infty} f_\mathcal{A}^{\QFA}(uv^n) \geq f_\mathcal{A}^{\QFA}(u)
        $,
        which implies
        $
            f_\mathcal{A}^{\QBA}(uv^\omega) \geq f_\mathcal{A}^{\QFA}(u)
        $.
    \end{proof}
    
    \subsection{Proof of Lemma \ref{lemma:op} (1)} \label{sec:proof-lemma-op-1}
    
    \begin{proof} [Proof of Lemma \ref{lemma:op} (1)]
        By Lemma \ref{lemma:lim} and Eq. (\ref{eq:qfa-direct-sum}), we have
    \[
    \begin{aligned}
    f_{a\mathcal{A}\oplus b\mathcal{B}}^{\QBA}(w)
    & = \limsup_{n\to\infty} f_{a\mathcal{A}\oplus b\mathcal{B}}^{\QFA}(w_n) \\
    & = \limsup_{n\to\infty} \rbra*{ \abs{a}^2 f_\mathcal{A}^{\QFA}(w_n) + \abs{b}^2 f_\mathcal{B}^{\QFA}(w_n) } \\
    & \leq \limsup_{n\to\infty} \abs{a}^2 f_\mathcal{A}^{\QFA}(w_n) 
    + \limsup_{n\to\infty} \abs{b}^2 f_\mathcal{B}^{\QFA}(w_n) \\
    & = \abs{a}^2 f_\mathcal{A}^{\QBA}(w) + \abs{b}^2 f_\mathcal{B}^{\QBA}(w).
    \end{aligned}
    \]
    On the other hand, we have:
    \[
    \begin{aligned}
    f_{a\mathcal{A}\oplus b\mathcal{B}}^{\QBA}(w)
    & = \limsup_{n\to\infty} \rbra*{ \abs{a}^2 f_\mathcal{A}^{\QFA}(w_n) + \abs{b}^2 f_\mathcal{B}^{\QFA}(w_n) } \\
    & \geq \max \bigg\{ \abs{a}^2 \limsup_{n\to\infty} f_\mathcal{A}^{\QFA}(w_n), 
     \abs{b}^2 \limsup_{n\to\infty} f_\mathcal{B}^{\QFA}(w_n) \bigg\} \\
    & = \max \left\{ \abs{a}^2 f_\mathcal{A}^{\QBA}(w), \abs{b}^2 f_\mathcal{B}^{\QBA}(w) \right\} \\
    & \geq \frac 1 2 \left( \abs{a}^2 f_\mathcal{A}^{\QBA}(w) + \abs{b}^2 f_\mathcal{B}^{\QBA}(w) \right) \\
    & \geq \frac 1 2 f_{a\mathcal{A}\oplus b\mathcal{B}}^{\QBA}(w).
    \end{aligned}
    \]
    Finally, the proof is obtained by combining the both inequalities. 
    
    For the special case that $\mathcal{A} = \mathcal{B}$, by Lemma \ref{lemma:lim} and Eq. (\ref{eq:qfa-direct-sum}), we have
    \begin{align*}
        f_{a\mathcal{A} \oplus b\mathcal{A}}^{\QBA}(w)
        & = \limsup_{n\to\infty} f_{a\mathcal{A}\oplus b\mathcal{A}}^{\QFA}(w_n) \\
        & = \limsup_{n\to\infty} f_{\mathcal{A}}^{\QFA}(w_n) \\
        & = f_{\mathcal{A}}^{\QBA}(w).
    \end{align*}
    \end{proof}
    
    \subsection{Proof of Lemma \ref{lemma:op} (2)} \label{sec:proof-lemma-op-2}

    \begin{proof} [Proof of Lemma \ref{lemma:op} (2)]
        By Lemma \ref{lemma:lim} and Eq. (\ref{eq:qfa-tensor-product}), we have
    \begin{align*}
        f_{\mathcal{A} \otimes \mathcal{B}}^{\QBA}(w)
        & = \limsup_{n\to\infty} f_{\mathcal{A}\otimes \mathcal{B}}^{\QFA}(w_n) \\
        & = \limsup_{n\to\infty} f_{\mathcal{A}}^{\QFA}(w_n) f_{\mathcal{B}}^{\QFA}(w_n) \\
        & \leq \limsup_{n\to\infty} f_{\mathcal{A}}^{\QFA}(w_n) \limsup_{n\to\infty} f_{\mathcal{B}}^{\QFA}(w_n) \\
        & = f_{\mathcal{A}}^{\QBA}(w) f_{\mathcal{B}}^{\QBA}(w).
    \end{align*}
    
    For the special case that $\mathcal{A} = \mathcal{B}$, by Lemma \ref{lemma:lim} and Eq. (\ref{eq:qfa-tensor-product}), we have
    \begin{align*}
        f_{\mathcal{A}^{\otimes k}}^{\QBA}(w)
        & = \limsup_{n\to\infty} f_{\mathcal{A}^{\otimes k}}^{\QFA}(w_n) \\
        & = \limsup_{n\to\infty} \left( f_{\mathcal{A}}^{\QFA}(w_n) \right)^k \\
        & = \left( \limsup_{n\to\infty} f_{\mathcal{A}}^{\QFA}(w_n) \right)^k \\
        & = \left( f_{\mathcal{A}}^{\QBA}(w) \right)^k.
    \end{align*}
    \end{proof}
    
    \subsection{Proof of Lemma \ref{lemma:op} (3)} \label{sec:proof-lemma-op-3}

    \begin{proof} [Proof of Lemma \ref{lemma:op} (3)]
        By Lemma \ref{lemma:lim} and Eq. (\ref{eq:qfa-ortho-complement}), we have
    \begin{align*}
         f_{\mathcal{A}}^{\QBA}(w) + f_{\mathcal{A}^\perp}^{\QBA}(w) 
         & = \limsup_{n\to\infty} f_{\mathcal{A}}^{\QFA}(w_n) + \limsup_{n\to\infty} f_{\mathcal{A}^\perp}^{\QFA}(w_n) \\
         & = \limsup_{n\to\infty} f_{\mathcal{A}}^{\QFA}(w_n) + \limsup_{n\to\infty} \left(1-f_{\mathcal{A}}^{\QFA}\right) \\
         & = 1 + \limsup_{n\to\infty} f_{\mathcal{A}}^{\QFA}(w_n)-\liminf_{n\to\infty} f_{\mathcal{A}}^{\QFA}(w_n) \\ 
         & \geq 1,
    \end{align*}
    and the equality holds if and only if $\limsup\limits_{n\to\infty} f_{\mathcal{A}}^\mathrm{MO}(w_n) = \liminf\limits_{n\to\infty} f_{\mathcal{A}}^\mathrm{MO}(w_n)$,
    which implies the existence of $\lim\limits_{n\to\infty} f_{\mathcal{A}}^\mathrm{MO}(w_n)$.
    \end{proof}
    
    \subsection{Proof of Lemma \ref{lemma:scaling-weighting}} \label{sec:proof-lemma-scaling-weighting}
    \begin{proof} [Proof of Lemma \ref{lemma:scaling-weighting}]
        Let $\mathcal{C}_\lambda = \rbra*{\mathcal{H}, \ket{s_0}, \Sigma, \{ U_\sigma : \sigma \in \Sigma \}, F }$ be a QBA for any $\lambda \in [0, 1]$, where:
        \begin{enumerate}
        \item $\mathcal{H} = \spanspace \cbra*{ \ket0, \ket1 }$,
        \item $F = \spanspace \cbra*{ \ket0 }$,
        \item $\ket{s_0} = \sqrt \lambda \ket 0 + \sqrt{1-\lambda} \ket 1$, and
        \item $U_\sigma = I$ for all $\sigma \in \Sigma$.
        \end{enumerate}
        It can be verified that $f_{\mathcal{C}_\lambda}^{\QBA}(w) = \lambda$ for every $w \in \Sigma^\omega$, and $f_{\mathcal{C}_\lambda}^{\QFA}(w) = \lambda$ for every $w \in \Sigma^*$.
        
        Let $\mathcal{B} = \mathcal{C}_\lambda \otimes \mathcal{A}$. Then, for every $w \in \Sigma^\omega$, we have
        \begin{align*}
            f_{\mathcal{B}}^{\QBA}(w)
            & = \limsup_{n\to\infty} f_{\mathcal{C}_\lambda \otimes \mathcal{A}}^{\QFA}(w_n) \\
            & = \limsup_{n\to\infty} f_{\mathcal{C}_\lambda}^{\QFA}(w_n) f_{\mathcal{A}}^{\QFA}(w_n) \\
            & = \limsup_{n\to\infty} \lambda f_{\mathcal{A}}^{\QFA}(w_n) \\
            & = \lambda f_{\mathcal{A}}^{\QBA}(w).
        \end{align*}
        
        Let $\mathcal{C} = \sqrt{\lambda} \mathcal{A} \oplus \sqrt{1-\lambda} \mathcal{C}_1$. Then, for every $w \in \Sigma^\omega$, we have
        \begin{align*}
        f_{\mathcal{C}}^{\QBA}(w)
        & = \limsup_{n\to\infty} f_{\sqrt{\lambda} \mathcal{A} \oplus \sqrt{1-\lambda} \mathcal{C}_1}^{\QFA}(w_n) \\
        & = \limsup_{n\to\infty} \left( \lambda f_{\mathcal{A}}^{\QFA}(w_n) + (1-\lambda) \right) \\
        & = \lambda f_{\mathcal{A}}^{\QBA}(w) + (1-\lambda). 
    \end{align*}
    \end{proof}
    
    \section{Pumping Lemmas for QBAs}
    
    \subsection{Proof of Theorem \ref{thm:pumping-lemma-helper}} \label{sec:proof-thm-pumping-lemma-helper}
    
    \begin{proof} [Proof of Theorem \ref{thm:pumping-lemma-helper}]
        Let $\mathcal{A} = (\mathcal{H}, \ket{s_0}, \Sigma, \{ U_\sigma : \sigma \in \Sigma \}, F )$ with $\mathcal{H}$ being $n$-dimensional.
    For any $w \in \Sigma^+$, by the Spectral Decomposition Theorem (see \cite{NC10}), there is a unitary matrix $V$ and a diagonal matrix $D$ such that $U_w = V^\dag DV$.
    Then by Lemma \ref{lemma:Dk-approx-for-pumping}, we choose $\delta = \varepsilon/4 > 0$, then there is a positive integer $k \leq \rbra*{c\delta}^{-n}$ for some constant $c > 0$ such that $D^k = I+\delta J$, where $J$ is a diagonal matrix with $\tr\rbra*{J^\dag J} < 1$.
    For any $u \in \Sigma^*$ and $v \in \Sigma^\omega$, we split the proof into two steps.
    
    \textbf{Step 1}. Let $\lambda$ be any real number such that $\lambda < f_\mathcal{A}^{\QBA}(uv)$, then by Lemma \ref{lemma:qba-alter-interpretation}, $\exists \varepsilon' > 0, \exists \ket\psi \in F, \exists \{ n_i \}, \forall i \in \mathbb{N}, \abs*{\braket{\psi}{s_{n_i}}}^2 > \lambda+\varepsilon'$, where $\ket{s_n}$ is the run on input $uv$ of $\mathcal{A}$. Without loss of generality, we may assume that $n_1 > \abs{u}$ (because $u$ is a finite word). Then $\ket{s_{n_i}} = U_{v_{n_i-\abs{u}}} U_u \ket{s_0}.$
        Now we consider the word $uw^kv$. Let $m_i = n_i + k\abs{w}$ and let $\ket{t_n}$ be the run on input $uw^kv$ of $\mathcal{A}$. Then
    \[
    \begin{aligned}
        \abs*{ \braket {\psi} {t_{m_i}} }^2 
        & = \abs*{ \bra{\psi} U_{v_{m_i-\abs{u}-k\abs{w}}} U_w^k U_u \ket{s_0} }^2 \\
        & = \abs*{ \bra{\psi} U_{v_{n_i-\abs{u}}} V^\dag D^k V U_u \ket{s_0} }^2 \\
        & = \abs*{ \bra{\psi} U_{v_{n_i-\abs{u}}} V^\dag (I+\delta J) V U_u \ket{s_0} }^2 \\
        & = \Big| \bra{\psi} U_{v_{n_i-\abs{u}}} U_u \ket{s_0}  + \delta \bra{\psi} U_{v_{n_i-\abs{u}}} V^\dag J V U_u \ket{s_0} \Big|^2 \\
        & = \abs*{ \braket{\psi}{s_{n_i}} + \delta \bra{\psi} U_{v_{n_i-\abs{u}}} V^\dag J V U_u \ket{s_0} }^2 \\
        & \geq \left( \abs*{\braket{\psi}{s_{n_i}}} - \delta \abs*{\bra{\psi} U_{v_{n_i-\abs{u}}} V^\dag J V U_u \ket{s_0}} \right)^2 \\
        & \geq \abs{\braket{\psi}{s_{n_i}}}^2 - 2\delta \\
        & > \lambda + \varepsilon' - 2\delta \\
        & > \lambda + \varepsilon' - \varepsilon.
    \end{aligned}
    \]
    By Lemma \ref{lemma:qba-alter-interpretation}, it holds that $f_\mathcal{A}^{\QBA}(uw^kv) > \lambda-\varepsilon.$ Since $\lambda$ can arbitrarily tend to $f_\mathcal{A}^{\QBA}(uv)$, we have:
    \begin{equation} \label{eq1:pumping-qba}
    f_\mathcal{A}^{\QBA}(uw^kv) \geq f_\mathcal{A}^{\QBA}(uv)-\varepsilon.
    \end{equation}

    \textbf{Step 2}. Let $\lambda$ be any real number such that $\lambda > f_\mathcal{A}^\mathrm{ND}(uv)$, then by Lemma \ref{lemma:qba-alter-interpretation}, there is a $\varepsilon' > 0$ such that for any state $\ket\psi \in F$, and any checkpoints $\{ n_i \}$, there is a $i \in \mathbb{N}$ such that $\abs*{\braket{\psi}{s_{n_i}}}^2 < \lambda - \varepsilon'$, where $\ket{s_n}$ is the run on $uv$ of $\mathcal{A}$. Without loss of generality, we may assume that $n_1 > \abs{u}$ (because $u$ is a finite word). Then $\ket{s_{n_i}} = U_{v_{n_i-\abs{u}}} U_u \ket{s_0}.$ Now we consider the word $uw^kv$. Let $m_i = n_i + k\abs{w}$ and $\ket{t_n}$ be the run on $uw^kv$ of $\mathcal{A}$. Then
    \[
      \begin{aligned}
        \abs*{ \braket {\psi} {t_{m_i}} }^2 
        & = \abs*{ \braket{\psi}{s_{n_i}} + \delta \bra{\psi} U_{v_{n_i-\abs{u}}} V^\dag J V U_u \ket{s_0} }^2 \\
        & \leq \left( \abs*{\braket{\psi}{s_{n_i}}} + \delta \abs*{\bra{\psi} U_{v_{n_i-\abs{u}}} V^\dag J V U_u \ket{s_0}} \right)^2 \\
        & \leq \abs*{\braket{\psi}{s_{n_i}}}^2 + 3\delta \\
        & < \lambda - \varepsilon_1 + 3\delta \\
        & < \lambda - \varepsilon' + \varepsilon.
      \end{aligned}
    \]
    By Lemma \ref{lemma:qba-alter-interpretation}, it holds that $f_\mathcal{A}^{\QBA}(uw^kv) < \lambda+\varepsilon.$ Since $\lambda$ can arbitrarily tend to $f_\mathcal{A}^{\QBA}(uv)$, we have:
      \begin{equation} \label{eq2:pumping-qba}
      f_\mathcal{A}^{\QBA}(uw^kv) \leq f_\mathcal{A}^{\QBA}(uv)+\varepsilon.
      \end{equation}

    Combining Eq. (\ref{eq1:pumping-qba}) and Eq. (\ref{eq2:pumping-qba}), we have
    \[
    \abs*{f_\mathcal{A}^{\QBA}(uv) - f_\mathcal{A}^{\QBA}(uw^kv)} \leq \varepsilon.
    \]
    Finally, we consider the relationship between $k$ and $\delta$. It holds that $k \leq (c\delta)^{-n} \leq (c\varepsilon/3)^{-n},$ and we complete the proof.
    \end{proof}
    
    \subsection{Proof of Theorem \ref{thm:pumping-lemma}} \label{sec:proof-thm-pumping-lemma}
    
    \begin{proof} [Proof of Theorem \ref{thm:pumping-lemma}]
        Let $L = \mathcal{L}_{>\lambda}^{\QBA}(\mathcal{A})$ for some QBA $\mathcal{A}$ and $\lambda \in [0, 1)$. 
        We first prove the first part. 
        By Theorem \ref{thm:pumping-lemma-helper}, for any $w \in \Sigma^+$ and any $\varepsilon > 0$, there are infinitely many $k$'s such that for any $u \in \Sigma^*$ and $v \in \Sigma^\omega$, $\abs*{f_\mathcal{A}^{\QBA}(uv)-f_\mathcal{A}^{\QBA}(uw^kv)} < \varepsilon$. If $uv \in L$, i.e., $f_\mathcal{A}^{\QBA}(uv) > \lambda$, then there is a $\varepsilon_0>0$ such that $f_\mathcal{A}^{\QBA}(uv) > \lambda+\varepsilon_0$. If we choose $\varepsilon = \varepsilon_0/2$, then
        \[
        f_\mathcal{A}^{\QBA}(uw^kv) > f_\mathcal{A}^{\QBA}(uv)-\varepsilon > \lambda+\varepsilon_0/2 > \lambda.
        \]
        As a result, we have $uw^kv \in \mathcal{L}_{>\lambda}^{\QBA}(\mathcal{A}) = L$.
        
        Then, we prove the second part. 
        If $v \in L$, i.e. $f_\mathcal{A}^{\QBA}(v) > \lambda$, then by Lemma \ref{lemma:alter}, we have: $\exists \varepsilon > 0, \exists \{ n_i \}, \forall i \in \mathbb{N}, \Abs*{P_F\ket{s_{n_i}}}^2 > \lambda+\varepsilon.$
    We choose the prefixes $x_i = v_{n_i}$, and thus
    $
        f_\mathcal{A}^{\QFA}(x_i) > \lambda
    $.
    Then for any $w \in \Sigma^+$, by Lemma \ref{lemma:self-rep}, we have:
    $
        f_\mathcal{A}^{\QBA}(x_i w^\omega) \geq f_\mathcal{A}^{\QFA}(x_i) > \lambda
    $,
    and thus $x_i w^\omega \in \mathcal{L}_{>\lambda}^{\QBA}(\mathcal{A}) = L$.
    \end{proof}
    
    \section{Non-Inclusion Relations of QBAs} \label{sec:proof-thm-almost-sure-notin-threshold}
    
    \begin{proof} [Proof of Theorem \ref{thm:almost-sure-notin-threshold}]
    We will show this non-inclusion result by a counterexample. 
    We choose a QBA $\mathcal{A} = \rbra*{ \mathcal{H}, \ket{s_0}, \Sigma, \{ U_\sigma : \sigma \in \Sigma \}, F }$, where:
    \begin{enumerate}
    \item $\mathcal{H} = \operatorname{span} \{ \ket0, \ket1 \}$,
    \item $\ket{s_0} = \ket0$,
    \item $\Sigma = \{ a, b \}$,
    \item $F = \operatorname{span} \{ \ket0 \}$, and
    \item $U_a = R_x\rbra*{\sqrt 2 \pi}$ and $U_b = R_x\rbra*{-\sqrt 2 \pi}$.
    \end{enumerate}
    Let $L = \mathcal{L}_{=1}^{\QBA}(\mathcal{A})$.
    We now use Theorem \ref{thm:pumping-lemma} (the second item) to show that $L \notin \mathbb{L}_{>\lambda}(\QBA)$ for any $\lambda \in [0, 1)$.
    We choose $v = a^\omega \in L$. For any prefix $x$ of $a^\omega$, say $x = a^n$ for some $n > 0$, and we choose $w = ab \in \Sigma^+$. Note that $xw^\omega = a^n(ab)^\omega \notin L$, and thus $L \notin \mathbb{L}^{>\lambda}({\QBA})$. As a result, we have: $\mathbb{L}_{=1}({\QBA}) \not \subseteq \mathbb{L}_{>\lambda}({\QBA})$ for any $\lambda \in [0, 1)$.

    It remains to show that $a^\omega \in L$ and $a^n(ab)^\omega \notin L$ for all $n > 0$.

    \textbf{Step 1}. Show that $a^\omega \in L$.
        
        By Lemma \ref{lemma:self-rep}, we obtain:
        $f_\mathcal{A}^{\QBA}(a^\omega) \geq f_\mathcal{A}^{\QFA}(\epsilon) = 1$,
        and thus $f_\mathcal{A}^{\QBA}(a^\omega) = 1$, i.e., $a^\omega \in L$.
        
    \textbf{Step 2}. Show that $a^n(ab)^\omega \notin L$.
        
        For any $n > 0$ and $k > 0$, we have
        \[
            \begin{aligned}
                f_\mathcal{A}^{\QFA} \left((a^n (ab)^\omega)_{n+2k}\right) & = \cos^2 \rbra*{ \frac {\sqrt2 n \pi} 2 }, \\
                f_\mathcal{A}^{\QFA} \left((a^n (ab)^\omega)_{n+2k+1}\right) & = \cos^2 \left( \frac {\sqrt2 (n+1) \pi} 2 \right).
            \end{aligned}
        \]
    By Proposition \ref{lemma:lim},
    \begin{align*}
        &f_\mathcal{A}^{\QBA}(a^n (ab)^\omega) =  \max \left\{ \cos^2 \left( \frac {\sqrt2 n \pi} 2 \right), \cos^2 \left( \frac {\sqrt2 (n+1) \pi} 2 \right) \right\} < 1.
    \end{align*}
    Therefore, $a^n(ab)^\omega \notin L$.
    
    \end{proof}
    
    \section{Pumping Lemma for \texorpdfstring{$\omega$}{ω}-Context-Free Languages} \label{app:wcfl-pumping}

    To prove Theorem \ref{thm:pumping-lemma-w-cfl}, we first recall the following pumping lemma for context-free languages from \cite{HMU06}.

    \begin{theorem} [Pumping lemma for $\CFL$, \cite{HMU06}] \label{thm:pumping-lemma-cfl}
        If $L \subseteq \Sigma^*$ is a context-free language, then there exists an integer $n \geq 1$ such that each word $z \in L$ with $\abs{z} \geq n$ can be written as $z = uvwxy$, where $u, v, w, x, y \in \Sigma^*$, such that
        \begin{itemize}
            \item[1.] $\abs{vwx} \leq n$,
            \item[2.] $\abs{vx} \geq 1$, and
            \item[3.] For all $k \in \mathbb{N}$, $uv^kwx^ky \in L$.
        \end{itemize}
    \end{theorem}
    
    Then, we will generalize the pumping lemma for context-free languages stated in Theorem \ref{thm:pumping-lemma-cfl} to that for $\omega$-context-free languages. 
    
    \begin{proof} [Proof of Theorem \ref{thm:pumping-lemma-w-cfl}]
        For any $\omega$-language $L \in \omega\mbox{-}\CFL$, every $\omega$-word in $L$ can be expressed as the concatenation of infinitely many finite context-free words by the $\omega$-Kleene closure (cf. \cite{CG77a,CG77b,Lin76}). That is, there are $m$ pairs of $\omega$-languages $U_i$ and $V_i$ such that $U_i, V_i \in \CFL$ and
        \[
            L = \bigcup_{i=1}^m U_i V_i^\omega.
        \]
        Let $z = \sigma_1 \sigma_2 \dots \in \Sigma^\omega$ be an $\omega$-word. If $z \in L$, then $z \in U_i V_i^\omega$ for some $1 \leq i \leq m$. Furthermore, $z$ can be written as $z = z_0 z_1 z_2 \dots$, where $z_0 \in U_i$ and $z_j \in V_i\setminus\{\epsilon\}$ for all $j \geq 1$. Note that each $V_i$ has its own ``pumping length'', say $n_i$ for all $1 \leq i \leq m$. That is, by Theorem \ref{thm:pumping-lemma-cfl}, for any $1 \leq i \leq m$, there is an integer $n_i \geq 0$ such that each $z \in V_i$ with $\abs{z} \geq n_i$ can be written as $z = uvwxy$ such that
        \begin{itemize}
            \item[1.] $\abs{vwx} \leq n_i$,
            \item[2.] $\abs{vx} \geq 1$, and
            \item[3.] For all $k \in \mathbb{N}$, $uv^kwx^ky \in V_i$.
        \end{itemize}

        Now let $n_0 = \max\limits_{1 \leq i \leq m} \{ n_i \}$ and $N_0 = 3n_0$. We choose $N_0$ to be the ``pumping length'' of $L$. Then, we consider the two cases as follows. 
        
        \begin{enumerate}
            \item There is a $j \geq 1$ such that $\abs{z_j} \geq n_0$. We apply Theorem \ref{thm:pumping-lemma-cfl} on $z_j$. Then, $z_j$ can be written as $z_j = abcde$, where $a, b, c, d, e \in \Sigma^*$, such that
            \begin{itemize}
            \item $\abs{bcd} \leq n_i \leq n_0$, $\abs{bd} \geq 1$, and
            \item $ab^kcd^ke \in V_i$ for all $k \in \mathbb{N}$.
            \end{itemize}
            Now we choose $u = z_0 z_1 \dots z_{j-1} a$, $v = b$, $w = c$, $x = d$, and $y = e z_{j+1} z_{j+2} \dots$, then it can be verified that
            \begin{enumerate}
            \item $\abs{vwx} = \abs{bcd} \leq n_0 < N_0$,
            \item $\abs{vx} = \abs{bd} \geq 1$, and
            \item for any $k \in \mathbb{N}$, we have
            $uv^kwx^ky = z_0 z_1 \dots z_{j-1} ab^kcd^ke z_{j+1} z_{j+2} \dots \in U_i V_i^\omega \subseteq L$.
            \end{enumerate}
            
            \item $\abs{z_j} < n_0$ for all $j \geq 1$. In this case, we choose $u = z_0$, $v = z_1$, $w = z_2$, $x = z_3$, and $y = z_4 z_5 \dots$, then it can be verified that
            \begin{itemize}
            \item $\abs{vwx} = \abs{z_1z_2z_3} < 3n_0 = N_0$,
            \item $\abs{vx} = \abs{z_1}+\abs{z_3} \geq 1$, and for all $k \in \mathbb{N}$, $uv^kwx^ky = z_0 z_1^k z_2 z_3^k z_4 \dots \in U_i V_i^\omega \subseteq L$.
            \end{itemize}
        \end{enumerate}
        From the above two cases, we conclude that $N_0$ is a valid ``pumping length'' of $L$, and these yield the proof.
    \end{proof}
    
    \section{Expressiveness of QBA beyond Classical} \label{sec:proof-thm-QBA-notin-RL}

    \begin{proof} [Proof of Theorem \ref{thm:QBA-notin-RL}]
        We first show the first part that $\mathbb{L}_{>\lambda}\rbra*{\QBA} \not \subseteq \omega\mbox{-}\RL$ for $\lambda \in \rbra*{0, 1}$. Let $\mathcal{A} = (\mathcal{H}, \ket{s_0}, \Sigma, \{ U_\sigma : \sigma \in \Sigma \}, F )$ be a QBA, where:
        \begin{enumerate}
        \item $\mathcal{H} = \spanspace \{ \ket0, \ket1 \}$,
        \item $\ket{s_0} = \ket0$,
        \item $\Sigma = \{ a, b \}$,
        \item $F = \spanspace \{ \ket0 \}$, and
        \item $U_a = R_x\rbra*{\sqrt 2 \pi}$ and $U_b = R_x\rbra*{-\sqrt 2 \pi}$.
        \end{enumerate}
        We choose $\lambda = 0.9$ and let $L = \mathcal{L}_{>\lambda}^{\QBA}(\mathcal{A})$. We use Theorem \ref{thm:pumping-lemma-w-reg} to show that $L$ is not $\omega$-regular.
        For any positive integer $n_0$, we choose the infinite word $w = a^{2n_0}b^{2n_0}(ab)^\omega \in L$ and choose $n = n_0$. Then for any split $w = xyz$, where $\abs{x} = n_0$ and $1 \leq \abs{y} \leq n_0$, we can find certain non-negative integer $k$, and then have $xy^kz = a^{2n_0+(k-1)\abs{y}}b^{2n_0}(ab)^\omega \notin L.$ By Theorem \ref{thm:pumping-lemma-w-reg}, we conclude that $L$ is not $\omega$-regular.

        It remains to show that: (1) for any $n \in \mathbb{N}$, $a^{2n}b^{2n}(ab)^\omega \in L$, and (2) for any $1 \leq m \leq n$, there is a non-negative integer $k$ such that $a^{2n+km}b^{2n}(ab)^\omega \notin L$.
        
        \begin{enumerate}
            \item $a^{2n}b^{2n}(ab)^\omega \in L$.
            
            For any $n \in \mathbb{N}$, note that the state of the run on input $a^{2n}b^{2n}(ab)^\omega$ of $\mathcal{A}$ at the checkpoint $n_i = 4n+2i$ is $\ket0 \in F$, and thus $f_\mathcal{A}^{\QBA}(a^{2n}b^{2n}(ab)^\omega) = 1 > 0.9 = \lambda$, i.e., $a^{2n}b^{2n}(ab)^\omega \in L$.
            
            \item For every $n \in \mathbb{N}$ and every $1 \leq m \leq n$, there exists a non-negative integer $k$ such that $a^{2n+km}b^{2n}(ab)^\omega \notin L$.
            
            For every $n \in \mathbb{N}$ and every $1 \leq m \leq n$, note that the state of the run on input $a^{2n+km}b^{2n}(ab)^\omega$ of $\mathcal{A}$ at the checkpoint $4n+km+2l$ and $4n+km+2l+1$ are $\ket*{s_{4n+km+2l}} = R_x\rbra*{\sqrt2 km\pi}\ket0$ and $\ket*{s_{4n+km+2l+1}} = R_x\rbra*{\sqrt2 (km+1)\pi}\ket0$, respectively, for all $l \in \mathbb{N}$. Thus
            \[
            \begin{aligned}
            f_\mathcal{A}^{\QBA}\rbra*{a^{2n+km}b^{2n}(ab)^\omega} 
            & = \max \cbra*{ \abs*{\braket{0}{s_{4n+km+2l}}}^2, \abs*{\braket{0}{s_{4n+km+2l+1}}}^2 } \\
            & = \max \cbra*{ \cos^2 \left( \frac {km\pi} {\sqrt2} \right), \cos^2 \left( \frac {(km+1)\pi} {\sqrt2} \right) } \\
            \end{aligned}
            \]
            Note that $\cos^2 (0.4) = 0.848353\dots < 0.9 = \lambda$ and $\cos^2(0.4+\frac{\pi}{\sqrt{2}}) = 0.752979\dots < 0.9 = \lambda.$ If we choose $k$ such that $\rbra*{{km\pi}/{\sqrt2}} \bmod 2\pi$ is close enough to $0.4$, then
            \[
            f_\mathcal{A}^{\QBA}\rbra*{a^{2n+km}b^{2n}(ab)^\omega} < \lambda,
            \]
            which implies that $a^{2n+km}b^{2n}(ab)^\omega \notin L$.
        \end{enumerate}
        Therefore, $L$ is not $\omega$-regular. As a result, we have $\mathbb{L}^{>\lambda}(\QBA) \not \subseteq \omega\mbox{-}\RL$.
        
        Next, we will show the second part that $\mathbb{L}_{=1}\rbra*{\QBA} \not \subseteq \omega\mbox{-}\CFL$. Let $\mathcal{A} = (H, \ket{s_0}, \Sigma, \{ U_\sigma : \sigma \in \Sigma \}, F )$ be a QBA, where:
        \begin{enumerate}
        \item $\mathcal{H} = \operatorname{span} \{ \ket0, \ket1 \}$,
        \item $\ket{s_0} = \ket0$,
        \item $\Sigma = \{ a, b, c \}$,
        \item $F = \operatorname{span} \{ \ket0 \}$, and
        \item $U_a = R_x\rbra{\rbra*{\sqrt 2+\sqrt 3} \pi}$, $U_b = R_x\rbra*{-\sqrt 2 \pi}$ and $U_c = R_x\rbra*{-\sqrt 3\pi}$.
        \end{enumerate}

        Let $L = \mathcal{L}_{=1}^{\QBA}(\mathcal{A})$.
        We use Theorem \ref{thm:pumping-lemma-w-cfl} to show that $L$ is not $\omega$-context-free.
        For any positive integer $n$, we choose $z = \rbra*{a^nb^nc^n}^\omega \in L$; for any split $z = uvwxy$ where $u, v, w, x \in \Sigma^*$ and $y \in \Sigma^\omega$ with $\abs{vwx} \leq n$ and $\abs{vx} \geq 1$, we can find certain non-negative integer $k$ such that $uv^kwx^ky \notin L$. Then, we conclude that $L \notin \omega\mbox{-}\CFL$.

        It remains to show that: (1) $\rbra*{a^nb^nc^n}^\omega \in L$, and (2) for every $n \in \mathbb{N}$ and every split $z = uvwxy$, there exists a non-negative integer $k$ such that $uv^kwx^ky \notin L$.
        
        \begin{enumerate}
            \item $\rbra*{a^nb^nc^n}^\omega \in L$ for every $n \geq 1$.
            
            For any $\rbra*{a^nb^nc^n}^\omega$ with $n\geq 1$, we choose the sequence of checkpoints $n_i = 3ni$. Note that $\ket{s_{n_i}} = \ket0$ for all $i \in \mathbb{N}$, where $\ket{s_i}$ is the run on input $\rbra*{a^nb^nc^n}^\omega$ of $\mathcal{A}$. Thus, $\abs*{\braket{0}{s_{n_i}}}^2 = 1$, and $f_\mathcal{A}^{\QBA}\rbra*{\rbra*{a^nb^nc^n}^\omega} = 1$, i.e., $\rbra*{a^nb^nc^n}^\omega \in L$.
            
            \item For every $n \in \mathbb{N}$ and every split $z = uvwxy$, there exists a non-negative integer $k$ such that $uv^kwx^ky \notin L$.
            
            For any split $z = uvwxy$ where $\abs{vwx} \leq n$ and $\abs{vx} \geq 1$, note that $vwx$ cannot contain all the letters in the alphabet $\Sigma$, which are $a$, $b$ and $c$. The analysis can be divided into several cases. Here, we only consider one of them, and the rest cases are similar. We consider the case that $vx$ contains $a$ but does not contain $b$. For convenience, we use $d(w)$ to denote the number of occurrences of symbol $d$ in word $w$, e.g. $a(aaab) = 3$. Then the case we are considering can be interpreted as $a(vx) \geq 1$ and $b(vx) = 0$.
            Note that for any prefix $p$ of $z$, we have $a(p) - b(p) \geq 0$ and the equality holds if and only if $\abs{p} \bmod 3n = 0$.
            We further note that there is a word $q \in \Sigma^*$, which is a prefix of $y$, such that $\abs{uvwxq} \bmod 3n = 0$, and thus $z = uvwxq(a^nb^nc^n)^\omega = uvwxqz.$ Then
            \begin{align*}
             a \rbra*{uv^kwx^kq} - b\rbra*{uv^kwx^kq} 
            & = a(uvwxq) - b(uvwxq) + (k-1)a(vx) \\
            & = (k-1)a(vx).
            \end{align*}
            On the other hand, for any $z \in \Sigma^\omega$, the run on input $z$ of $\mathcal{A}$ is
            \begin{align*}
            \ket*{s_i} = R_x \Big( \sqrt2 & \rbra*{a(z_i)-b(z_i)}\pi 
            + \sqrt3 \rbra*{a(z_i)-c(z_i)} \Big) \ket0
            \end{align*}
            for every $i \in \mathbb{N}$.
            Now we consider the run $\ket{s_n}$ on input $uv^kwx^ky$ of $\mathcal{A}$, and the checkpoint $n_i = \abs{uv^kwx^kq}+i$ for all $i \in \mathbb{N}$. Note that
            \[
            \begin{aligned}
            \ket{s_{n_i}}
            & = U_{z_i} R_x \Big( \sqrt2(a(uv^kwx^kq)-b(uv^kwx^kq))\pi 
             + \sqrt3(a(uv^kwx^kq)-c(uv^kwx^kq)) \Big) \ket0 \\
            & = U_{z_i} R_x \Big( \sqrt2((k-1)a(vx))\pi + 
            \sqrt3(a(uv^kwx^kq)-c(uv^kwx^kq)) \Big) \ket0 \\
            & = R_x \Big( \sqrt2((k-1)a(vx)+a(z_i)-b(z_i))\pi  + \sqrt3(a(uv^kwx^kq)-c(uv^kwx^kq)  +a(z_i)-c(z_i)) \Big) \ket0.
            \end{aligned}
            \]
            We further note that if we choose $k=2$, then 
            \begin{align*}
            (k-1)a(vx)+a(z_i)-b(z_i) 
            \geq (k-1)a(vx)
            = a(vx) \geq 1.
            \end{align*}
            Therefore, $\ket{s_{n_i}} \neq \ket0$ for any $i \in \mathbb{N}$.
            Moreover, note that $\ket{s_{n_i}}$ has a cycle of length $3n$, i.e., $\ket*{s_{n_i}} = \ket*{s_{n_{i+3n}}}$. If we choose
            \[
            M = \max_{0 \leq i < 3n} \cbra*{ \abs{\braket{0}{s_{n_i}}}^2 } < 1,
            \]
            then $\abs*{\braket{0}{s_{n_i}}}^2 \leq M < 1$ for all $i \in \mathbb{N}$. By Proposition \ref{prop:def-qba-by-limsup}, we have:
            \begin{align*}
                f_\mathcal{A}^{\QBA}\rbra*{uv^2wx^2y} & = \limsup_{n \to \infty} \abs*{\braket{0}{s_n}}^2 \\
                & \leq \max\cbra*{\abs*{\braket{0}{s_n}}^2} \\
                & = M < 1,
            \end{align*}
            i.e. $uv^2wx^2y \notin L$.
        \end{enumerate}
        Hence, $L$ is not $\omega$-context-free, which implies that $\mathbb{L}_{=1}\rbra*{\QBA} \not \subseteq \omega\mbox{-}\CFL$.
    \end{proof}
    
    \section{Proof of Theorem \ref{thm:closure-union}} \label{sec:thm-closure-union}

    \begin{proof} [Proof of Theorem \ref{thm:closure-union}]
        We first show the first part that $\mathbb{L}_{>0}\rbra*{\QBA}$ is closed under union.
        Suppose $\mathcal{A}_1$ and $\mathcal{A}_2$ are the QBAs of $L_1$ and $L_2$, respectively. That is, $L_1 = \mathcal{L}_{>0}^{\QBA}\rbra*{\mathcal{A}_1}$ and $L_2 = \mathcal{L}_{>0}^{\QBA}\rbra*{\mathcal{A}_2}$.
        Let $\mathcal{A}_{12} = \frac 1 {\sqrt2} \mathcal{A}_1 \oplus \frac 1 {\sqrt2} \mathcal{A}_2$.
        We will prove that $\mathcal{L}_{>0}^{\QBA}\rbra*{\mathcal{A}_{12}} = L_1 \cup L_2$ as follows.
        
        \begin{enumerate}
            \item $\mathcal{L}_{>0}^{\QBA}\rbra*{\mathcal{A}_{12}} \subseteq L_1 \cup L_2$.
            
            For any $w \in \mathcal{L}_{>0}^{\QBA}\rbra*{\mathcal{A}_{12}}$, i.e., $f_{\mathcal{A}_{12}}^{\QBA}(w) > 0$,
            by Lemma \ref{lemma:op} (1), we have
            \[
                \max \cbra*{ f_{\mathcal{A}_{1}}^{\QBA}(w), f_{\mathcal{A}_{2}}^{\QBA}(w) } \geq f_{\mathcal{A}_{12}}^{\QBA}(w) > 0.
             \]
            Then, either $f_{\mathcal{A}_{1}}^{\QBA}(w) > 0$ or $f_{\mathcal{A}_{2}}^{\QBA}(w) > 0$, which implies $w \in L_1 \cup L_2$. Thus, $\mathcal{L}_{>0}^{\QBA}\rbra*{\mathcal{A}_{12}} \subseteq L_1 \cup L_2$.
            
            \item $L_1 \cup L_2 \subseteq \mathcal{L}_{>0}^{\QBA}\rbra*{\mathcal{A}_{12}}$.
            
            For any $w \in L_1 \cup L_2$, i.e. 
            \[
            \max \cbra*{ f_{\mathcal{A}_{1}}^{\QBA}(w), f_{\mathcal{A}_{2}}^{\QBA}(w) } > 0,
            \]
            by Lemma \ref{lemma:op} (1), we have
            \[
                f_{\mathcal{A}_{12}}^{\QBA}(w) \geq \frac 1 2 \max \cbra*{ f_{\mathcal{A}_{1}}^{\QBA}(w), f_{\mathcal{A}_{2}}^{\QBA}(w) } > 0.
            \]
            Then $w \in \mathcal{L}_{>0}^{\QBA}\rbra*{\mathcal{A}_{12}}$. Thus, $L_1 \cup L_2 \subseteq \mathcal{L}_{>0}^{\QBA}\rbra*{\mathcal{A}_{12}}$.
        \end{enumerate}
        The above arguments together show that $\mathbb{L}_{>0}\rbra*{\QBA}$ is closed under union.
        
        Next, we show the second part $\mathbb{L}_{>\lambda}\rbra*{\QBA}$ is closed under union in the limit for $\lambda \in \rbra*{0, 1}$. Let $\mathcal{A}$ and $\mathcal{B}$ be two QBAs such that $L_A = \mathcal{L}_{>\lambda}^{\QBA}(\mathcal{A})$ and $L_B = \mathcal{L}_{>\lambda}^{\QBA}(\mathcal{B})$.
        For every $k \geq 1$, we define
        \[
            \mathcal{M}_k = \frac 1 {\sqrt2} \left( \mathcal{A}^{\otimes k} \oplus \mathcal{B}^{\otimes k} \right),
        \]
        and $L_k = \mathcal{L}_{>\lambda^k}^{\QBA}(\mathcal{M}_k)$. By Theorem \ref{thm:relation-threshold-semantics}, we know that $L_k \in \mathbb{L}_{>\lambda^{k}}(\QBA) = \mathbb{L}_{>\lambda}(\QBA)$. In the following, we will show that $\lim\limits_{k \to \infty} L_k = L_A \cup L_B$. We first show that $L_k$ is monotonic, then show that $L_A \cup L_B$ is an upper bound of all $L_k$, and finally show that this upper bound is tight.

        \textbf{Claim 1}. For any $1 \leq k < m$, we have $L_k \subseteq L_m$.
        
        \begin{proof}
            Let $w \in L_k$. By Lemma \ref{lemma:alter}, we have
            \[
                f_{\mathcal{M}_k}^{\QBA}(w) = \limsup_{n\to\infty} f_{\mathcal{M}_k}^{\QFA}(w_n) > \lambda^k.
            \]
            By Eq. (\ref{eq:qfa-direct-sum}), we have
            \[
            \begin{aligned}
            \left( f_{\mathcal{M}_k}^{\QFA}(w_n) \right)^{1/k} 
            & = \left( \frac 1 2 \left( \rbra*{f_\mathcal{A}^{\QFA}(w_n)}^k + \rbra*{f_\mathcal{B}^{\QFA}(w_n)}^k \right) \right)^{1/k} \\
            & \leq \left( \frac 1 2 \left( \rbra*{f_\mathcal{A}^{\QFA}(w_n)}^m + \rbra*{f_\mathcal{B}^{\QFA}(w_n)}^m \right) \right)^{1/m} \\
            & = \left( f_{\mathcal{M}_m}^{\QFA}(w_n) \right)^{1/m}.
            \end{aligned}
            \]
            Taking the supremum limit on both sides, we have
            \begin{align*}
            \limsup_{n\to\infty} \left( f_{\mathcal{M}_m}^{\QFA}(w_n) \right)^{1/m}
            & \geq \limsup_{n\to\infty} \left( f_{\mathcal{M}_k}^{\QFA}(w_n) \right)^{1/k} \\
            & = \left( \limsup_{n\to\infty} f_{\mathcal{M}_k}^{\QFA}(w_n) \right)^{1/k} \\
            & > \lambda.
            \end{align*}
            By Lemma \ref{lemma:alter} again, we have
            \[
                f_{\mathcal{M}_m}^{\QBA}(w) = \limsup_{n\to\infty} f_{\mathcal{M}_m}^{\QFA}(w_n) > \lambda^m,
            \]
            which means $w \in L_m$. Therefore, $L_k \subseteq L_m$.
        \end{proof}

    \textbf{Claim 2}. For any $k \in \mathbb{N}$, $L_k \subseteq L_A \cup L_B$.
    
    \begin{proof}
        Let $w \in L_k$. By Lemma \ref{lemma:alter}, we have
        \[
            f_{\mathcal{M}_k}^{\QBA}(w) = \limsup_{n\to\infty} f_{\mathcal{M}_k}^{\QFA}(w_n) > \lambda^k.
        \]
        On the other hand, if $w \notin L_A$ and $w \notin L_B$, i.e., $f_{\mathcal{A}}^{\QBA}\rbra*{w} \leq \lambda$ and $f_{\mathcal{B}}^{\QBA}\rbra*{w} \leq \lambda$, then by Eq. (\ref{eq:qfa-direct-sum}), we have
        \[
        \begin{aligned}
        \limsup_{n\to\infty}  f_{\mathcal{M}_k}^{\QFA}(w_n) 
        & = \limsup_{n\to\infty} \frac 1 2 \left( \rbra*{f_\mathcal{A}^{\QFA}(w_n)}^k + \rbra*{f_\mathcal{B}^{\QFA}(w_n)}^k \right) \\
        & \leq \frac 1 2 \left( \limsup_{n\to\infty} \rbra*{f_\mathcal{A}^{\QFA}(w_n)}^k + \limsup_{n\to\infty} \rbra*{f_\mathcal{B}^{\QFA}(w_n)}^k \right) \\
        & = \frac 1 2 \left( \rbra*{f_\mathcal{A}^{\QBA}(w)}^k + \rbra*{f_\mathcal{B}^{\QBA}(w)}^k \right)
         < \lambda^k.
        \end{aligned}
        \]
        A contradiction arises. As a result, $w \in L_A$ or $w \in L_B$, i.e., $w \in L_A \cup L_B$, which leads to $L_k \subseteq L_A \cup L_B$.
    \end{proof}
    
    The above two claims imply that the sequence $\cbra*{ L_k }$ of $\omega$-languages has a limit, and
    \[
    \lim_{k \to \infty} L_k \subseteq L_A \cup L_B.
    \]
    Then, we will show that this inclusion is not proper. 

    \textbf{Claim 3}. For any proper subset $L \subset L_A \cup L_B$, there is a $k \geq 1$ such that $L_k \not \subseteq L$. 
    
    \begin{proof}
        For any proper subset $L \subset L_A \cup L_B$, there is an $\omega$-word $w \in L_A \cup L_B$ but $w \notin L$. Without loss of generality, we may assume that $w \in L_A$, i.e., $f_{\mathcal{A}}^{\QBA}\rbra*{w} > \lambda$. By Lemma \ref{lemma:alter}, there is a sequence $\cbra*{n_i}$ of checkpoints such that $f_{\mathcal{A}}^{\QFA} \rbra*{w_{n_i}} > \lambda$. Note that
        \[
            \lim_{k \to \infty} \left( \frac {f_\mathcal{A}^{\QFA}\rbra*{w_{n_i}}} {\lambda} \right)^k = +\infty.
        \]
        Then, there is an integer $k \geq 1$ such that 
        \[
        \rbra*{ \frac {f_\mathcal{A}^{\QFA}\rbra*{w_{n_i}}} {\lambda} }^k > 4.
        \]
        Then, we have
        \begin{align*}
            \frac 1 2 \left( \rbra*{f_\mathcal{A}^{\QFA}(w_{n_i})}^k + \rbra*{f_\mathcal{B}^{\QFA}(w_{n_i})}^k \right) 
            \geq \frac 1 2 \rbra*{f_\mathcal{A}^{\QFA}(w_{n_i})}^k > 2 \lambda^k.
        \end{align*}
        Finally, by Lemma \ref{lemma:alter} and Eq. (\ref{eq:qfa-direct-sum}), we have 
        \begin{align*}
            f_{\mathcal{M}_k}^{\QBA}(w) = \limsup_{n\to\infty} \frac 1 2 \left( \rbra*{f_\mathcal{A}^{\QFA}(w_{n})}^k + \rbra*{f_\mathcal{B}^{\QFA}(w_{n})}^k \right) 
            \geq 2 \lambda^k > \lambda^k,
        \end{align*}
        i.e., $w \in L_k$. This means that we find a $k \geq 1$ such that $L_k \notin L$. 
    \end{proof}
    
    Through the above claims, we conclude that
    $
    \lim\limits_{k \to \infty} L_k = L_A \cup L_B
    $,
    and we complete the proof.
    \end{proof}
    
    \section{Supplementary Proof of Theorem \ref{thm:non-closure}} \label{app:supp-thm-non-closure}
    
    In the proof the Theorem \ref{thm:non-closure} that $\mathbb{L}_{>\lambda}\rbra*{\QBA}$ is not closed under limits for $\lambda \in (0, 1)$, $\lim\limits_{k \to \infty} L_k = \mathcal{L}_{=1}^{\QBA}(\mathcal{A})$ is left unproven. Here, we fill this gap. 
    
    First, it can be seen that $L_k$ is monotonic. Specifically, for every $1 \leq k < m$, we have $L_k \supseteq L_m$. This shows the existence of $\lim\limits_{k \to \infty} L_k$. Moreover, it can be shown that $\mathcal{L}_{=1}^{\QBA}(\mathcal{A})$ is a lower bound of every $L_k$, i.e., $\mathcal{L}_{=1}^{\QBA}(\mathcal{A}) \subseteq L_k$ for every $k \geq 1$. The two claims can be obtained by similar arguments in Theorem \ref{thm:closure-union}. Then, these together yield that
    \[
        \lim\limits_{k \to \infty} L_k \supseteq \mathcal{L}_{=1}^{\QBA}(\mathcal{A}).
    \]
    To show the equality, suppose $\lim\limits_{k \to \infty} L_k = L \supset \mathcal{L}_{=1}^{\QBA}(\mathcal{A})$. Then, there is an $\omega$-word such that $w \in L$ but $w \notin \mathcal{L}_{=1}^{\QBA}(\mathcal{A})$, i.e., $f_{\mathcal{A}}^{\QBA}\rbra*{w} < 1$. Now let
    \[
        k' = \ceil*{\frac{2}{1-f_{\mathcal{A}}^{\QBA}\rbra*{w}}} \geq 1.
    \]
    It can be verified that 
    \[
        f_\mathcal{A}^{\QBA}\rbra*{w} < 1 - \frac{1}{k'+10},
    \]
    and thus $w \notin L_{k'}$. A contradiction arises. Finally, we conclude that $\lim\limits_{k \to \infty} L_k = \mathcal{L}_{=1}^{\QBA}(\mathcal{A})$.
    
\end{document}